\documentclass[11pt,a4paper,epsf,epsfig,psfrag]{article}
\usepackage{jheppub}
\usepackage{amsmath,graphicx,amsfonts, mathrsfs,amssymb}
\usepackage{amsmath,epsfig}
\usepackage{amssymb,amsfonts}
\usepackage{color}
\usepackage{latexsym}
\usepackage{epsfig}
\usepackage{pdfsync}
\usepackage{marvosym}
\newbox\pippobox

\title{Critical exponents of finite temperature chiral phase transition in soft-wall AdS/QCD models}

\author[a]{Jianwei Chen}
\author[b,c]{Song He}
\author[d,e,f]{Mei Huang}
\author[a]{Danning Li}

\affiliation[a]{Department of Physics and Siyuan Laboratory, Jinan University, Guangzhou 510632, P.R. China}
\affiliation[b]{Max Planck Institute for Gravitational Physics (Albert Einstein Institute) Am M\"{u}hlenberg 1, 14476 Golm, Germany}
\affiliation[c]{Center for Theoretical Physics and College of Physics, Jilin University, Changchun, P. R. China}
\affiliation[d]{Institute of High Energy Physics, Chinese Academy of Sciences, Beijing 100049, P.R. China}
\affiliation[e]{University of Chinese Academy of Sciences, Beijing 100049, P.R. China}
\affiliation[f]{Theoretical Physics Center for Science Facilities, Chinese Academy of Sciences, Beijing 100049, P.R. China}

\emailAdd{chenjianwei@stu2016.jnu.edu.cn}
\emailAdd{hesong17@gmail.com}
\emailAdd{huangm@ihep.ac.cn}
\emailAdd{lidanning@jnu.edu.cn}

\abstract{Criticality of chiral phase transition at finite temperature is investigated in a soft-wall AdS/QCD model with $SU_L(N_f)\times SU_R(N_f)$ symmetry, especially for $N_f=2,3$ and $N_f=2+1$. It is shown that in quark mass plane($m_{u/d}-m_s$) chiral phase transition is second order at a certain critical line, by which the whole plane is divided into first order and crossover regions. The critical exponents $\beta$ and $\delta$, describing critical behavior of chiral condensate along temperature axis and light quark mass axis,  are extracted both numerically and analytically. The model gives the critical exponents of the values $\beta=\frac{1}{2}, \delta=3$ and $\beta=\frac{1}{3}, \delta=3$ for $N_f=2$ and $N_f=3$ respectively.  For $N_f=2+1$, in small strange quark mass($m_s$) region, the phase transitions for strange quark and $u/d$ quarks are strongly coupled, and the critical exponents are $\beta=\frac{1}{3},\delta=3$; when $m_s$ is larger than $m_{s,t}=0.290\rm{GeV}$, the dynamics of light flavors($u,d$) and strange quarks decoupled and the critical exponents for $\bar{u}u$ and $\bar{d}d$ becomes $\beta=\frac{1}{2},\delta=3$, exactly the same as $N_f=2$ result and the mean field result of 3D Ising model; between the two segments, there is a tri-critical point at $m_{s,t}=0.290\rm{GeV}$, at which $\beta=0.250,\delta=4.975$. In some sense, the current results is still at mean field level, and we also showed the possibility to go beyond mean field approximation by including the higher power of scalar potential and the temperature dependence of dilaton field, which might be reasonable in a full back-reaction model. The current study might also provide reasonable constraints on constructing a realistic holographic QCD model, which could describe both chiral dynamics and glue-dynamics correctly.}

\begin{document}
\maketitle
\section{Introduction}
\label{sec-int}

Spontaneous chiral symmetry breaking, as well as color confinement, characterize the vacuum of Quantum Chromodynamics(QCD), which is considered as fundamental theory of the strong force. It is believed that temperature effect could drive a transition between the chirally asymmetric phase at low temperature and a chirally symmetric phase at high temperature. Understanding the nature of this phase transition(usually called 'chiral phase transition'), as well as deconfinement phase transition,  are of great importance from theoretical, phenomenological and experimental aspects \cite{chiral-th,Hatsuda:1985eb,Laermann:2003cv,Schafer:1996wv,Stephanov:1998dy,Karsch:2000kv,chiral-rhic,Palmese:2016rtq,Schukraft:2015dna}.

\begin{figure}[h]
\begin{center}
\epsfxsize=7.5 cm \epsfysize=7.5 cm \epsfbox{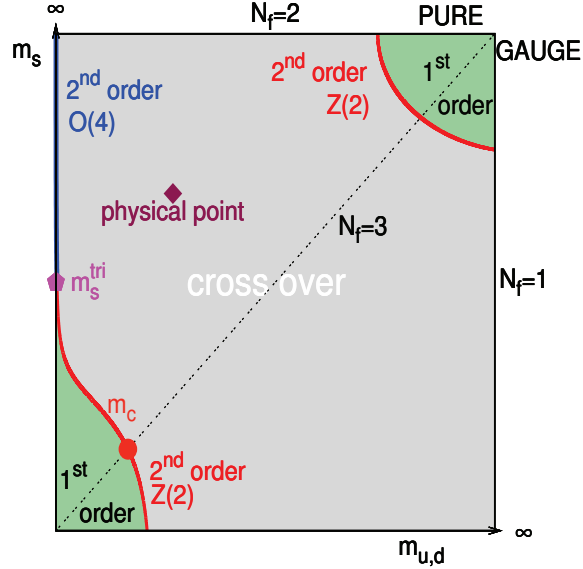} \
\end{center}
\caption{The expected phase diagram in the quark mass plane $m_{u/d}-m_s$ space(Taken from \cite{Ding:2015ona}).} \label{columbia-plot}
\end{figure}

As pointed out in \cite{chiral-th}, QCD phase transition might depend on the number of flavors $N_f$ and masses of quarks($m_u, m_d$ and $m_s$). Based on theoretical analysis, model calculations and lattice simulations, the expected phase structure in quark mass plane is summarized in the so called 'Columbia plot'\cite{columbia-plot-origin}, as shown in Fig.\ref{columbia-plot}(taken from \cite{Ding:2015ona}, see also \cite{columbia-plot-figure,Endrodi:2007gc} for other similar forms). In this sketch plot, both in the upper right and lower left corners, around the three-flavor chiral limit($m_{u,d,s}=0$) and pure gauge($m_{u,d,s}=\infty$) limit respectively, the phase transitions are of first order. In the intermediate part, the transition is widely accepted to be a continuous one, usually called 'crossover'. The boundaries between the first order regions and the crossover region are second order lines(the blue and red solid lines in Fig.\ref{columbia-plot}), at which the system become critical. Since the long range infrared fluctuations dominate at the critical point, only the dimensionality and relevant symmetries play important roles in the critical behavior while the microscopic details of the system are less relevant(For more details, please refer to \cite{Fisher:1974uq,Hohenberg:1977ym}). Understanding the critical properties and seeking the location of the second order lines are quite meaningful, since they would affect the location\cite{Fodor:2001pe,Fodor:2004nz} or even the existence\cite{deForcrand:2006pv} of the critical endpoint on the QCD phase diagram(in $T-\mu$ plane), which is one of the essential goals of Relativistic Heavy Ion Collisions\cite{Aggarwal:2010cw,Odyniec:2013aaa,Luo:2017faz}.

Generally, the second order lines are distinguished by the universality classes, described by critical behavior of the thermodynamical quantities or the corresponding order parameters. Near the small quark mass corner, chiral symmetry in Lagrangian level is a good approximation and the transition is depicted well by chiral condensate(also called 'quark condensate') $\sigma\equiv\langle\bar{\psi}\psi\rangle$. Chiral symmetry is broken and restored when $\sigma\neq0$ and $\sigma=0$ respectively. Near the critical point, the scaling behavior of chiral condensate are represented by critical exponents($\beta,\delta$)\footnote{In this work, we omit $\alpha,\gamma$ related to $C\simeq t^\alpha , \chi_m\equiv \frac{\partial\sigma}{\partial m}\simeq t^{\gamma}$, since in the current model the calculation of specific heat $C$ and susceptibility $\chi_m$ with finite quark mass requires the regularization of free energy, which is out of the scope of this paper.} along different axis in the following way:
\begin{eqnarray}
\sigma \simeq t^\beta,    \sigma  \simeq (m-m_c)^{1/\delta},
\label{eqcritical}
\end{eqnarray}
with $t=\frac{T_c-T}{T_c}$ and $T_c, m_c$ the critical temperature and quark mass respectively. From mean field calculation, in which the correlation of fluctuations are neglected, one gets $\beta=\frac{1}{2}, \delta=3$ in three dimension, which is not in agreement with experimental result \cite{Pelissetto:2000ek}.

Many efforts have been made on going beyond mean field approximation for QCD phase transition, e.g., from lattice QCD (LQCD)simulations\cite{AliKhan:2000wou,Ejiri:2009ac,Karsch:2010ya,Kaczmarek:2011zz,Burger:2011zc}, functional renormalization group(FRG)\cite{Grahl:2014fna,Wang:2015bky}, Dyson-Schwinger equations(DSEs)\cite{Fischer:2011pk,Fischer:2012vc}, $\epsilon$ expansion \cite{chiral-th,Yee:2017sir} and so on. Despite of the possibilities of first order transition in two-flavor chiral limit due to $U_A(1)$ restoration\cite{DElia:2005nmv,Chandrasekharan:2007up,Cossu:2013uua,Tomiya:2014mma}, it is summarized that when $m_{u/d}=0$ and $m_s$ is sufficiently large(the solid blue line in Fig.\ref{columbia-plot}), the second order line belongs to $O(4)$ class, with $\beta\approx 0.385,  \delta \approx 4.824$ \cite{Kanaya:1994qe,Engels:2001bq,Engels:2003nq}; while when $m_{u,d,s}$ is not very large(the solid red line in the lower-left corner) it belongs to $Z(2)$ classes, with $\beta\approx 0.327,  \delta \approx 4.789$\cite{Campostrini:2002cf,Bazavov:2017xul,Ding:2017giu}. Between the $Z(2)$ and $O(4)$ segments, there might be a tri-critical point. The relative locations of the tri-critical point and physical quark masses would affect the properties of critical end point in $T-\mu$ plane, when considering the curvature of the critical surface at $\mu_B=0$\cite{Ding:2015ona,deForcrand:2006pv,Philipsen:2007rj} together.



Generally speaking, lattice simulations are the most reliable first principle method dealing with strong interactions. However, when the quark mass is small, lattice is expensive, and it is still hard to control when chemical potential is large. So it is better to investigate this issue from different kinds of method and try to summarize reliable information from varies of ways. Besides the traditional methods, the AdS/CFT correspondence\cite{Maldacena:1997re,Gubser:1998bc,Witten:1998qj}(See also reviews\cite{Aharony:1999ti,Erdmenger:2007cm,deTeramond:2012rt,Adams:2012th}) provides a new approach for understanding non-perturbative physics of QCD.  For critical phenomena in bottom-up holographic QCD, it is shown that in Einstein-Maxwell-Dilaton model the critical exponents of thermodynamical quantities near the critical end pont are still in the mean field level\cite{DeWolfe:2011ts,DeWolfe:2010he,Finazzo:2016psx,Knaute:2017lll,Chen:2018vty}, due to the suppression of large $N$ effect.  However, from a more general Gauge/Gravity duality point of view and going out of the limitation of large $N$ phenomenologically, if one believes that the holographic approach could describe the strong correlation system, one might expect better description of critical phenomena, which is governed by the strong correlations instead of the interaction details. Besides, the critical exponents from the order parameter, chiral condensate, have  not been discussed in those models. Therefore, it is still quite interesting to study critical behavior of chiral phase transition, and try to see whether there are clues to go beyond mean field approximation.

In bottom-up holographic approach, chiral phase transition has been studied by several groups \cite{Iatrakis:2010jb,Jarvinen:2011qe,Alho:2012mh,Alho:2013hsa,Colangelo:2011sr,Dudal:2015wfn,Evans:2016jzo,Chelabi:2015cwn,Chelabi:2015gpc,Fang:2015ytf,Li:2016gfn,Li:2016smq,Bartz:2016ufc,Fang:2016nfj,Gursoy:2016ofp,Bartz:2017jku,Fang:2018vkp} . In the soft-wall AdS/QCD model\cite{Karch:2006pv}, which gives a good description of both chiral symmetry breaking and hadron spectra, the structure of chiral critical line in quark mass plane agrees very well with the 'Columbia plot' in Fig.\ref{columbia-plot}\cite{Li:2016smq,Bartz:2017jku,Fang:2018vkp}. The critical scaling behavior of chiral condensate along the axis of external field has been studied in a bottom-up holographic model\cite{Evans:2010np}. However, the critical exponents $\beta$ and $\delta$ of chiral critical point along $T$ and $m_q$ axis, as defined in Eq.(\ref{eqcritical}), has not yet been analyzed. Accordingly, considering the theoretical and phenomenological importance of critical phenomena of QCD, we will present an analysis on the critical scaling of chiral phase transition and try to extract the critical exponents in soft-wall AdS/QCD models.

The paper is organized as follows. We will firstly give a brief introduction about soft-wall AdS/QCD models in Sec.\ref{sec-soft}. Then in Sec.\ref{sec-twof}, we will analyze the critical behavior of two-flavor soft-wall model. We will give a constraint on the construction of dilaton field from chiral phase transition. We will also show the possibilities to go beyond mean field approximation in two-flavor case. Then, in Sec.\ref{sec-threef}, we will turn to three-flavor case with an additional t'hooft determinant interaction term representing the instanton effect. In Sec.\ref{sec-twoplusonef}, we will analyze the $N_f=2+1$ cases, with $m_u=m_d\neq m_s$. Finally, a short summary is given in Sec.\ref{sec-sum}. Moreover, due to the compactness of description, we leave some technical details on extracting the critical exponents analytically to the Appendix.

\section{Chiral dynamics in soft-wall models}
\label{sec-soft}

As mentioned in the introduction part, the soft-wall model\cite{Karch:2006pv} offers a good start point to describe linear confinement and chiral symmetry breaking, which are the two most important features of low energy QCD. Its extended models \cite{Gherghetta-Kapusta-Kelley,Gherghetta-Kapusta-Kelley-2,YLWu,YLWu-1,Cui:2013xva,Li:2012ay,Li:2012im,Li:2013oda,Vega:2016gip,Capossoli:2015ywa,Capossoli:2016kcr,Capossoli:2016ydo,Zollner:2017nnh,He:2013qq} could describe hadron physics in good agreement with experimental data. The action of soft-wall model takes the form\cite{Karch:2006pv}
\begin{eqnarray}\label{action}
 S=&&-\int d^5x
 \sqrt{-g}e^{-\Phi}Tr(D_M X^+ D^M X+V_X(|X|)).
\end{eqnarray}
Here, since we will only focus on describing chiral phase transition at finite temperature, we neglect the part related to gauge field in the original soft-wall model, which is expected to be vanished at zero chemical potential background. In the above action, $X$ is the $N_f \times N_f$ matrix-valued scalar field, $\Phi$ is the dilaton field, $V_X(|X|)$ is the scalar potentail, $M$ is the space time index taking values from $0,1,...,4$ and $g$ is the determinant of the metric $g_{MN}$. The above action is invariant under $SU_L(N_f)\times SU_R(N_f)$ gauge transformation \footnote{Of course, it should be considered in the full form with gauge filed.} . In this work, we will mainly consider $N_f=2,3$ along with $N_f=2+1$($N_f=3$ with $m_u=m_d\neq m_s$), and take $V_X=M_5^2 X^{+}X+\lambda|X|^4+\gamma Re[det(X)]$, with $M_5^2=-3$ (we will always take the AdS radius $L=1$ in this work) from the AdS/CFT prescription $M_5^2=(\Delta-p)(\Delta+p-4)$\cite{Witten:1998qj} by taking $\Delta=3, p=0$  . The coefficient $\gamma$ of the t'hooft determinant term will reduce to mass term in two-flavor case, and we will only consider this term in three-flavor case. The term $\lambda|X|^4$ has been shown to be important for the meson spectra and spontaneous chiral symmetry breaking, and we will take positive value of $\lambda$.

The scalar field $X^{\alpha,\beta}$ is supposed to be dual to the operator $\bar{\psi}^\alpha \psi^\alpha$, with $\psi$ the quark field and $\alpha$ the flavor index. In QCD vacuum, only the diagonal components of this operator has non-zero expectation value, which is just the chiral condensate. Therefore, in holographic framework, we assume the diagonal structure of the solution of $X$, i.e. $X=\text{Diag}\{\chi_{u},\chi_{d},...\}$. Phenomenologically, the configuration of the dilaton filed should be fixed by the experimental data of meson spectra and the requirement of dynamical breaking of chiral symmetry. In the following study, we will follow our previous study\cite{Chelabi:2015gpc} and take
\begin{eqnarray}\label{int-dilaton}
\Phi(z)=-\mu_1^2z^2+(\mu_1^2+\mu_0^2)z^2\tanh(\mu_2^2z^2),
\end{eqnarray}
which approaches $-\mu_1^2 z^2$ when $z\rightarrow0$ and $\mu_0^2z^2$ when $z\rightarrow\infty$, satisfying the requirement of dynamical chiral symmetry breaking and linear confinement respectively. As in\cite{Chelabi:2015gpc}, the free parameters are fixed to be
\begin{eqnarray}
\mu_0=0.430 \rm{GeV},\mu_1=0.830\rm{GeV}, \mu_2=0.176\rm{GeV}.\label{muvalues}
\end{eqnarray}
In view of the symmetry in 4D theory at finite temperature, the metric ansatz would be taken as
\begin{eqnarray}
dS^2=g_{MN}dx^{M}dx^{N}=e^{2A(z)}(-f(z)dt^2+\frac{1}{f(z)}dz^2+dx_idx^i).
\end{eqnarray}
Strictly, $A, f, \Phi$ could be solved self-consistently from certain kinds of gravity system(like what has been done in \cite{Li:2012ay,Li:2013oda,Li:2011hp,Cai:2012xh}). Here, for simplicity, we will consider the AdS-Schwarzchild(AdS-SW) black hole solutions
\begin{eqnarray}
A(z)&=&-\log(z),\label{As}\\
f(z)&=&1-\frac{z^4}{z_h^4}\label{f}.
\end{eqnarray}
In the above expression,  $z_h$ is the horizon where $f(z)=0$, and it is related to the temperature $T$ by formula
\begin{eqnarray}
T=\frac{1}{\pi z_h}.\label{Tzh}
\end{eqnarray}

Generally, one can start from three-flavor case and  assume
\begin{eqnarray}
X=\left(
    \begin{array}{ccc}
      \frac{\chi_u(z)}{\sqrt{2}} & 0 & 0 \\
      0 & \frac{\chi_d(z)}{\sqrt{2}} & 0 \\
      0 & 0 &\frac{\chi_s(z)}{\sqrt{2}}\\
    \end{array}
  \right).
\end{eqnarray}
If $m_u\neq m_d \neq m_s$, one expects $\chi_u\neq\chi_d\neq\chi_s$. Then the action Eq.(\ref{action}) reduces effectively to
\begin{eqnarray}\label{eff-action}
S[\chi_u,\chi_d,\chi_s]=-\int d^5x
 \sqrt{-g}e^{-\Phi}\left\{\Sigma_{i=u,d,s}\left[\frac{g^{zz}}{2}\chi_i^{'2}-\frac{3}{2}\chi_i^2+v_4\chi_i^4\right]+3v_3\chi_u\chi_d\chi_s\right\}.
\end{eqnarray}
Here $\chi^{'}$ denotes the derivative with respective to $z$, and $v_3\equiv\frac{\gamma}{6\sqrt{2}},v_4\equiv\frac{\lambda}{4}$ are redefinition of $\lambda,\gamma$.
Then one can derive the equation of motion for $\chi_u,\chi_d,\chi_s$ as follows.
\begin{eqnarray}
\chi_u^{''}+(3A_s^{'}-\Phi^{'}+\frac{f^{'}}{f})\chi_u^{'}+\frac{e^{2A_s}}{f}(3\chi_u-3v_3\chi_d\chi_s-4v_4\chi_u^3)&=&0,\label{eom-chiu}\\
\chi_d^{''}+(3A_s^{'}-\Phi^{'}+\frac{f^{'}}{f})\chi_d^{'}+\frac{e^{2A_s}}{f}(3\chi_d-3v_3\chi_u\chi_s-4v_4\chi_d^3)&=&0,\label{eom-chid}\\
\chi_s^{''}+(3A_s^{'}-\Phi^{'}+\frac{f^{'}}{f})\chi_s^{'}+\frac{e^{2A_s}}{f}(3\chi_s-3v_3\chi_u\chi_d-4v_4\chi_s^3)&=&0.\label{eom-chis0}
\end{eqnarray}
Considering only two degenerate flavors with $m_u=m_d, \chi_s=0$ and taking $v_3=0$, one has $\chi_u=\chi_d\equiv\chi$ and reaches
\begin{eqnarray}
\chi^{''}+(3A_s^{'}-\Phi^{'}+\frac{f^{'}}{f})\chi^{'}+\frac{e^{2A_s}}{f}(3\chi-4v_4\chi^3)=0.\label{eom-chi2}
\end{eqnarray}
Similarly, if considering three degenerate flavors, one has $\chi_u=\chi_d=\chi_s\equiv\chi$ and gets
\begin{eqnarray}
\chi^{''}+(3A_s^{'}-\Phi^{'}+\frac{f^{'}}{f})\chi^{'}+\frac{e^{2A_s}}{f}(3\chi-3v_3\chi^2-4v_4\chi^3)=0.\label{eom-chi3}
\end{eqnarray}
Moreover, if considering two degenerate light flavos plus one heavier flavor, then one has $\chi_u=\chi_d\equiv\chi_l\neq\chi_s$ and the above equations reduce to
\begin{eqnarray}
\chi_l^{''}+(3A_s^{'}-\Phi^{'}+\frac{f^{'}}{f})\chi_l^{'}+\frac{e^{2A_s}}{f}(3\chi_l-3v_3\chi_l\chi_s-4v_4\chi_l^3)&=&0,\label{eom-chil}\\
\chi_s^{''}+(3A_s^{'}-\Phi^{'}+\frac{f^{'}}{f})\chi_s^{'}+\frac{e^{2A_s}}{f}(3\chi_s-3v_3\chi_l^2-4v_4\chi_s^3)&=&0.\label{eom-chis}
\end{eqnarray}

The asymptotic solution near the boundary $z=0$ could be extracted perturbatively order by order as
\begin{eqnarray}
\chi_l&=&c_l z-3 c_l c_s v_3 z^2-(\mu_1^2-2c_l^2v_4+\frac{9}{2}c_s^2v_3^2+\frac{9}{2}c_l^2v_3^2)c_lz^3\log( z)+d_l z^3+...,\label{chiluv}\\
\chi_s&=&c_s z-3 c_l^2 v_3 z^2-(\mu_1^2-2c_s^2v_4-9 c_l^2v_3^2)c_s z^3\log(z)+d_s z^3+... .\label{chisuv}
\end{eqnarray}
Here, $c_l, d_l, c_s, d_s$ are four integral constants at ultraviolet boundary of the two coupled second order ordinary derivative equations(ODEs) Eqs.(\ref{eom-chil},\ref{eom-chis}). Since $X$ is supposed to be dual to chiral condensate, one should map the integral constants to the sources and the operators ($m_l,m_s,\sigma_l\equiv\langle\bar{u}u(\bar{d}d)\rangle,\sigma_s\equiv\langle\bar{s}s\rangle$) as
\cite{Karch:2006pv}
\begin{eqnarray}
c_l&=&m_l \zeta,\\
d_l&=&\frac{\sigma_l}{ \zeta},\\
c_s&=&m_s \zeta,\\
d_s&=&\frac{\sigma_s}{\zeta},
\end{eqnarray}
with $\zeta=\frac{\sqrt{3}}{2\pi}$ fixed by matching the two point correlation function of $\langle\bar{\psi}\psi(q)\bar{\psi}\psi(0)\rangle$ to the 4D calculation\cite{Cherman:2008eh}. In the calculation, the quark mass $m_l, m_s$ should be considered as a physical input. Then one needs two additional boundary conditions in order to solve the two second order ODEs. Fortunately, we found that there are two natural conditions that $\chi_l,\chi_s$ should be regular everywhere, especially at horizon
\begin{eqnarray}
|\chi_l(z_h)|<\infty,|\chi_s(z_h)|<\infty.\label{chiir}
\end{eqnarray}
Input the quark masses and the above boundary conditions, one can use a shooting method and solve $\sigma_l, \sigma_s$ out of the equations  of motion Eqs.(\ref{eom-chil},\ref{eom-chis}). The procedure for solving two degenerate and three degenerate flavor could be simply got by taking $m_l=m_s\equiv m$ and $\sigma_l=\sigma_s\equiv \sigma$, and we will not repeat here. Based on the above preparation, one can investigate the temperature and quark mass dependent behavior of chiral condensate from soft-wall AdS/QCD models. For the progress in this direction, please refer to \cite{Dudal:2015wfn,Chelabi:2015cwn,Chelabi:2015gpc,Fang:2015ytf,Li:2016gfn,Li:2016smq,Bartz:2016ufc,Fang:2016nfj,Bartz:2017jku,Fang:2018vkp}. In the following sections,  we will focus on investigating the critical behavior of chiral condensate in different situations.

\section{Critical exponents with two degenerate light quarks}
\label{sec-twof}

\begin{figure}[h]
\begin{center}
\epsfxsize=6.5 cm \epsfysize=6.5 cm \epsfbox{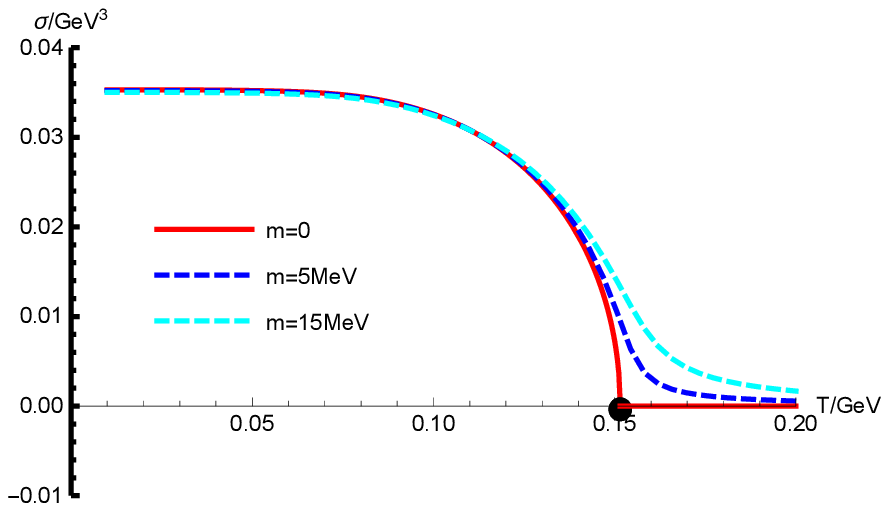}
\hspace*{0.1cm} \epsfxsize=6.5 cm \epsfysize=6.5 cm
\epsfbox{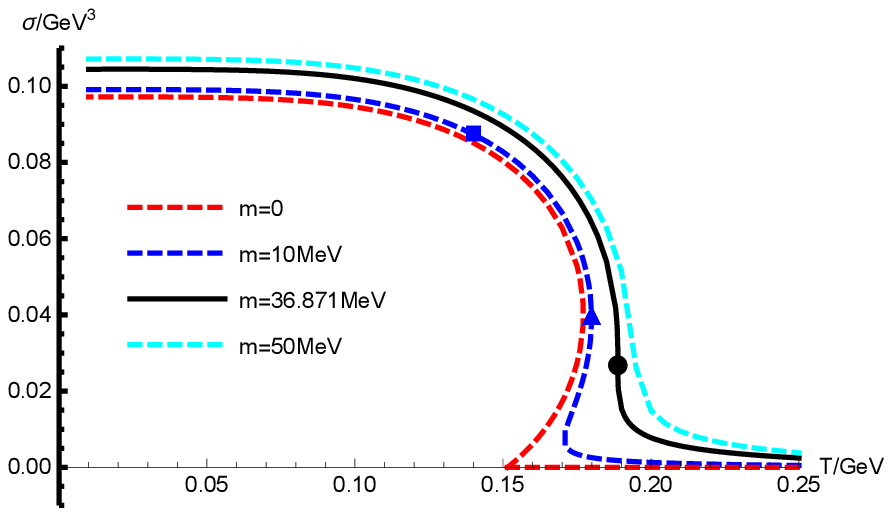} \vskip -0.05cm \hskip 0.15 cm
\textbf{( a ) } \hskip 6.5 cm \textbf{( b )} \\
\end{center}
\caption{Chiral condensate $\sigma$ as a function of temperature $T$ for cases $N_f=2$ and $N_f=3$. The corresponding dilaton profile are shown in Eqs.(\ref{int-dilaton}-\ref{muvalues}). The metric is taken as the AdS-Schwarzchild black hole solution as shown in Eqs.(\ref{As},\ref{f}). Panel.(a) and (b) are results solved from Eq.(\ref{eom-chi2}) and Eq.(\ref{eom-chi3}) respectively. In Panel.(a), the results of $m=0,5,15{\rm MeV}$ for two degenerate quarks with $v_3=0, v_4=8$ are shown in red-solid, blue-dashed, cyan-dashed lines respectively. The black dot denotes the critical point of the second order phase transition at $T_{c,2}=0.1515...{\rm MeV}, m_{c,2}=0, \sigma_{c,2}=0$. In Panel.(b), the results of $m=0$,$10$,$37$,$50\rm{MeV}$ for three degenerate quarks with $v_3=-3, v_4=8$ are shown in red-dashed, blue-dashed, black-solid, cyan-dashed lines respectively. The black dot denotes the critical point of the second order phase transition at $T_{c,3}=0.1888...\rm{GeV}$, $m_{c,3}=0.036871...\rm{GeV}$, $\sigma_{c,3}=0.02724...\rm{GeV}^3$, at which $\frac{d\sigma}{dT}=\infty$. The blue rectangle and triangle dots are labeled for the description in Appendix.\ref{ana-three}.      }
\label{sigma-T-degenerate}
\end{figure}

Firstly, we will consider the case with two degenerate light quarks in the model with dilaton configuration defined in Eqs.(\ref{int-dilaton},\ref{muvalues}) and scalar interaction with $v_3=0,v_4=-8$. The equations of motion for this case are shown in Eq.(\ref{eom-chi2}). The temperature and quark mass dependent behavior of chiral condensate have been carefully studied in\cite{Chelabi:2015cwn,Chelabi:2015gpc}. For the convenience of description, part of the results are displayed again in Fig.\ref{sigma-T-degenerate}(a).

\begin{figure}[h]
\begin{center}
\epsfxsize=6.5 cm \epsfysize=6.5 cm \epsfbox{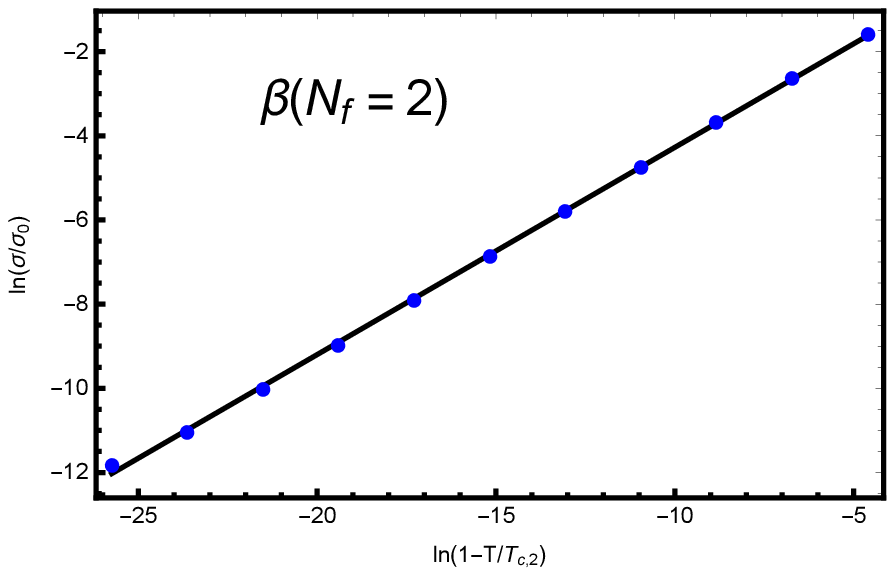}
\hspace*{0.1cm} \epsfxsize=6.5 cm \epsfysize=6.5 cm
\epsfbox{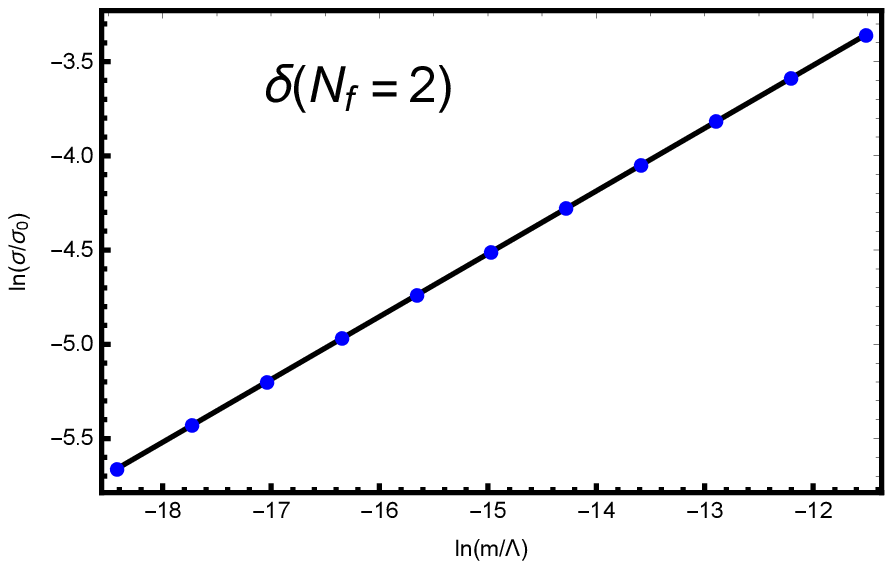} \vskip -0.05cm \hskip 0.15 cm
\textbf{( a ) } \hskip 6.5 cm \textbf{( b )} \\
\end{center}
\caption{Temperature and quark mass dependence of chiral condensate with two degenerate quarks near the critical point $T_{c,2}=0.1515...{\rm MeV}, m_{c,2}=0, \sigma_{c,2}=0$ . The corresponding dilaton profile are shown in Eqs.(\ref{int-dilaton}-\ref{muvalues}) and the scalar potential parameters are taken as $v_3=0,v_4=8$. Panel.(a) gives $\ln(\frac{\sigma}{\sigma_0})$ as a function of $\ln(t)$, with fixed quark mass $m=m_{c,2}=0$. The blue dots are model calculation and the black solid straight line are linear fitting of the dots, which has the form $y=0.648+0.492 x$.  Panel.(b) gives $\ln(\frac{\sigma}{\sigma_0})$ as a function of $\ln(\frac{m}{\Lambda})$, with $\Lambda=1\rm{GeV}$ a parameter introduced to make the quantity inside the logarithmic function dimensionless.  The blue dots are model calculation and the black solid straight line are linear fitting of the dots, which has the form $y=0.480+0.333 x$. }
\label{ces-twof}
\end{figure}

In Fig.\ref{sigma-T-degenerate}(a), we give the results of $m=0,5,15{\rm MeV}$ for two degenerate quarks with $v_3=0, v_4=8$ in red-solid, blue-dashed, cyan-dashed lines respectively.  From the figure, we could see that there are transitions between the low temperature chirally asymmetric phase with chiral condensate around $\sigma_0=0.035\rm{GeV}^3$ and the high temperature chirally symmetric phase with almost vanished chiral condensation. Here, in a strict sense, chiral symmetry is restored when $\sigma$ is exactly zero. However, since finite quark mass would break chiral symmetry of QCD Lagrangian explicitly, one might consider the small value tails in Fig.\ref{sigma-T-degenerate}(a) as effects of explicit breaking from small quark mass. In this sense, we also consider the high temperature tails as symmetry restored phase.  It is easy to see that in chiral limit with $m=0$(the red solid line in Fig.\ref{sigma-T-degenerate}(a)) the phase transition is a second order transition, while it turns to be a crossover one at any finite quark mass($m>0$). The critical point(the black dot in Fig.\ref{sigma-T-degenerate}(a)) of the second order transition locates at $T_{c,2}=0.1515...{\rm MeV}, m_{c,2}=0, \sigma_{c,2}=0$\footnote{In order to determine the critical exponents, one has to determine the critical point to very high accuracy, but we will not show the digits all here.}. According to the definition of critical exponent $\beta$  in Eq.(\ref{eqcritical}), we  tune the temperature $T$ slightly below $T_{c,2}$ and solve the corresponding chiral condensate with  fixing quark mass $m=m_{c,2}=0$. From Eq.(\ref{eqcritical}), the critical behavior of chiral condensate near $T_{c,2}$ obeys certain critical scaling $\sigma=\sigma-\sigma_{c,2}\propto(T_{c,2}-T)^\beta$. Then one expects that $\ln(\sigma-\sigma_{c,2})=\ln(\sigma)$ will depend linearly on $\ln(T_{c,2}-T)$ near $T_{c,2}$. The slope of the linear line is just the critical exponent $\beta$. Since $\sigma, T$ are not  dimensionless quantities, we rescale them with $\sigma_0, T_{c,2}$ and consider $\ln(\frac{\sigma}{\sigma_0})$ as a function of $\ln(t)\equiv\ln(1-\frac{T}{T_{c,2}})$. It is easy to understand that this rescaling would not change the critical exponent. Therefore, we plot $\ln(\frac{\sigma}{\sigma_0})$ as a function of $\ln(t)$  from $\ln(t)=-25$ to $\ln(t)=-5$ in Fig.\ref{ces-twof}(a). From the plot, we find that all the data points(the blue dots in Fig.\ref{ces-twof}(a)) lie perfectly in a straight line. From a best fitting, the straight line is of the form $y=0.648+0.492x$ with the slope $0.492$. This fact shows that  numerically $\beta=0.492$, which is very close to the value $\beta=\frac{1}{2}$ from 3D mean field calculation.

Similarly, the critical exponent $\delta$ could be extracted based on Eq.(\ref{eqcritical}). We tune quark mass $m$ slightly larger than critical value $m_{c,2}=0$ and solve $\sigma$ with fixing temperature $T=T_{c,2}$. It is expected that $\sigma=\sigma-\sigma_{c,2}\propto (m-m_{c,2})^{1/\delta}=m^{1/\delta}$ near the critical point. Equivalently, $\ln(\frac{\sigma}{\sigma_0})$ should depends on $\ln(\frac{m}{\Lambda})$. Here, since $m_{c,2}=0$, we introduced a mass scale $\Lambda$ to keep the quantity inside the logarithmic function dimensionless. Since the exact value of $\Lambda$ will not affect the slope, we simply take it to be $\Lambda=1\rm{GeV}$. Then we plot $\ln(\frac{\sigma}{\sigma_0})$ as a function of $\ln(\frac{m}{\Lambda})$ from $\ln(\frac{m}{\Lambda})\approx-18$ to $\ln(\frac{m}{\Lambda})\approx-11$ in  Fig.\ref{ces-twof}(b).  It is quite obvious that the numerical data points of  $(\ln(\frac{m}{\Lambda}),\ln(\frac{\sigma}{\Lambda}))$(the blue dots in Fig.\ref{ces-twof}(b)) lie perfectly in a straight line, which reveals the critical scaling. From a best linear fitting, the straight line has the form $0.480+0.333x$. From our definition, the slope of this straight line is just the inverse of  critical exponent $\delta$. So, numerically, we get $\delta\approx\frac{1}{0.333}=3.003$, also very close to the value $\delta=3$ from 3D mean field calculation.

The numerical analysis in the above reveals $\beta=0.492, \delta=3.003$. This result is so close to the 3D mean filed result $\beta=\frac{1}{2},\delta=3$ that one might suspect that the difference comes just from numerical errors. Usually, one can not study the critical scaling exactly at the critical temperature, which might be the largest source of the numerical errors. Hence, to be more accurate, we try to investigate the critical scaling behavior analytically. Note that the critical point locates at $T_{c,2}=0.1515...{\rm MeV}, m_{c,2}=0, \sigma_{c,2}=0$, where both quark mass and chiral condensate are zero. The critical solution of $\chi(z)$ vanishes also. Thus, near the critical point, $\chi(z)$ should be very small and could be expand according to the small deviation from the critical point $\delta T\equiv T_{c,2}-T$ and $\delta m\equiv m-m_{c,2}$. For later convenience, we can make a coordinate transformation $z= z_h s=\frac{s}{\pi T}$ to Eq.(\ref{eom-chi2}), under which the horizon would be set to $s=1$. Then Eq.(\ref{eom-chi2}) becomes
\begin{eqnarray}
\ddot{\chi}(s)-(\frac{3}{s}+\frac{4s^3}{1-s^4}+\dot{\phi}_{T}(s))\dot{\chi}(s)- \frac{1}{s^2(1-s^4)}(3-4v_4\chi^3)=0.\label{eom-chi2-s0}
\end{eqnarray}
Here $\phi_T(s)\equiv \Phi(\frac{s}{\pi T})$ is the dilaton in new coordinate, and the 'dot' in $\dot{\phi}_{T}(s)$ representing derivative with respect to $s$. Unlike the original form $\Phi(z)$, the apparent form of $\phi_T(s)$ will depend on $T$, so we label it with a lower index $T$. The expansion of $\chi$ near the critical point would be
\begin{eqnarray}
\chi=\chi_{m_c,1}\delta T^{\epsilon_1}+\chi_{m_c,2} \delta T^{\epsilon_2}+...
\end{eqnarray}
under small $T$ deviation $\delta T$ together with fixing $m=m_{c,2}=0$, and
\begin{eqnarray}
\chi=\chi_{T_c,1} \delta m^{\kappa_1}+\chi_{T_c,2} \delta m^{\kappa_2}+...
\end{eqnarray}
under small $m$ deviation $\delta m$ together with fixing $T=T_{c,2}$. Without loss of generality, we assume that the power $\epsilon_{i}$ and $\kappa_i$ satisfy $0<\epsilon_{i}<\epsilon_{i+1}$ and $0<\kappa_{i}<\kappa_{i+1}$. The functions $\chi_{m_c,i},\chi_{T_c,i}$ should be considered as functions of $s$ with fixing $m=m_{c,2}$ and $T=T_{c,2}$.  Inserting the expansion into Eq.(\ref{eom-chi2-s0}), one can obtain the equation for $\chi_{m_c,i}$ and $\chi_{T_c,i}$  order by order perturbatively. Studying the properties of the solution $\chi_{m_c,i}$ and $\chi_{T_c,i}$, it is possible to get the information of $\epsilon_i$ and $\kappa_i$. Since the analytical derivation is quite technical, we leave it to Appendix.\ref{ana-two} and only discuss the main physical result here(For details, please refer to Appendix.\ref{ana-two}).

Firstly, we go to the leading expansion with respect to $\delta T$. It is not hard to get
\begin{eqnarray}\label{eq-chi0-se2}
\ddot{\chi}_{m_c,1}-(\frac{3}{s}+\frac{4s^3}{1-s^4}+\dot{\phi}_{T_c})\dot{\chi}_{m_c,1}+ \frac{3}{s^2(1-s^4)}\chi_{m_c,1}=0.
\end{eqnarray}
All the functions above should be considered as functions of $s$. Since we are expanding around the critical point, we have taken $T=T_{c,2}$ in $\phi_T$. The other condition $m=m_{c,2}=0$ are reflected in the boundary condition of $\chi_{m_c,1}$. Note that the boundary $z=0$ is transformed to $s=0$. The boundary expansion in $z$ coordinate $\chi(z)=m_q \zeta z+...$ becomes $\frac{m \zeta}{\pi T} s+...$ . So, keeping $m=m_{c,2}=0$ means keeping $\dot{\chi}_0\equiv\dot{\chi}_{m_c,1}(s=0)=0$. In another side, the regularity of $\chi$ requires the leading expansion taking the form $\chi_{m_c,1}(s)=c_0(1+\frac{3}{4}(s-1)+o((1-s)))$.   Actually, all these conditions give us a very simple criteria on whether the dilaton profile could describe the spontaneous symmetry breaking or not.  Given that Eq.(\ref{eq-chi0-se2}) is linear, we can simply take $c_0=1$ and solve it. Supposing that now we are dealing with a general dilaton profile $\phi_T(s)$. If $\phi_T(s)$ could give a second order phase transition in chiral limit from $\sigma\neq0$ to $\sigma =0$. Then, there should be a certain value of $T$, at which the solution of $\chi_{m_c,1}(s)$ satisfying both $\chi_{m_c,1}(s)=c_0(1+\frac{3}{4}(s-1)+o((1-s)))$ and $\dot{\chi}_0=0$. One can solve Eq.(\ref{eq-chi0-se2}), replacing $\dot{\phi}_{T_c}$ with $\dot{\phi}_{T}$ and taking $c_0=1$. After getting the solution, one can extract $\dot{\chi}_0$ . If there is a certain $T_1$ giving $\dot{\chi}_0$, then the dilaton profile can give second order phase transition in chiral limit and $T_1$ is just the transition temperature.

\begin{figure}[h]
\begin{center}
\epsfxsize=6.5 cm \epsfysize=6.5 cm \epsfbox{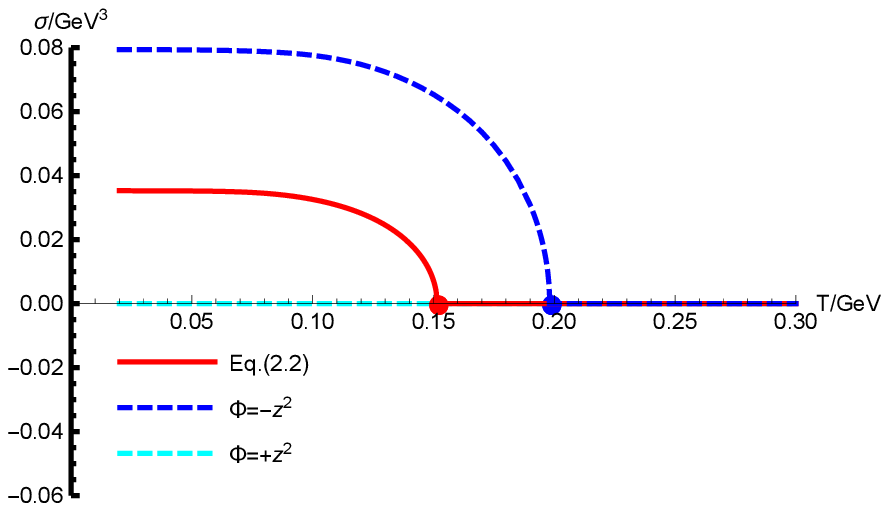}
\hspace*{0.1cm} \epsfxsize=6.5 cm \epsfysize=6.5 cm \epsfbox{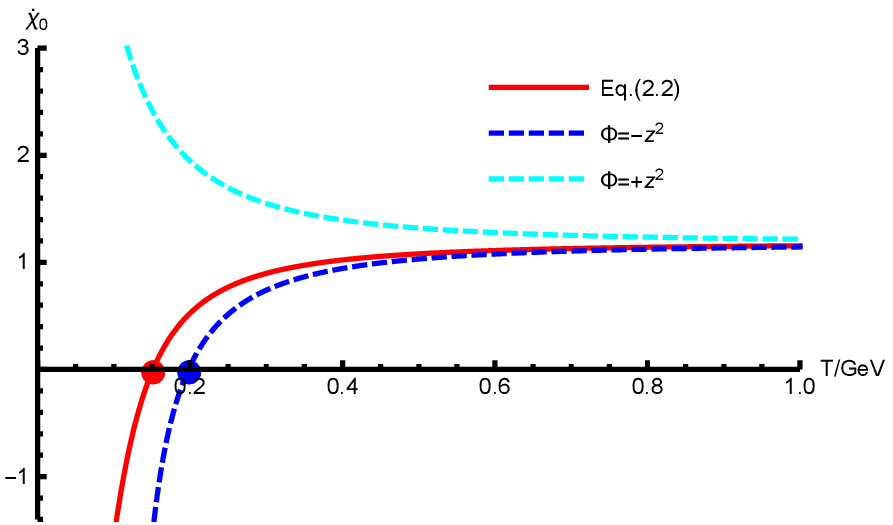}
\vskip -0.05cm \hskip 0.15 cm
\textbf{( a ) } \hskip 6.5 cm \textbf{( b )} \\
\end{center}
\caption{Temperature dependent chiral condensate(Panel.(a)) and the corresponding solution of $\dot{\chi}_{0}$(Panel.(b)),  with dilaton in Eq.(\ref{int-dilaton})(the red solid lines), $\Phi=-z^2$(the blue dashed lines), $\Phi=z^2$(the cyan dashed lines). The scalar potential takes the form $v_3=0, v_4=8$. The red and blue dots represent the transition temperature, which locates at $T=0.1515...\rm{GeV}$ and $T=0.1980...\rm{GeV}$ for dilaton in Eq.(\ref{int-dilaton}) and $\Phi=-z^2$ respectively.  The high temperature limit of all the three lines are the same and approach $-\frac{\Gamma^2(\frac{1}{4})}{2\sqrt{\pi}(\psi^{(0)}(\frac{1}{4})-\psi^{(0)}(\frac{3}{4}))}\approx1.18$, with $\psi^{(n)}(z)$ the $n^{th}$ derivative of the digamma function $\psi(z)\equiv\frac{\Gamma^{'}(z)}{\Gamma(z)}$. }
\label{check-order}
\end{figure}

In order to make this point clearer, we take three well studied dilaton configurations (see Ref.\cite{Chelabi:2015gpc}), $\Phi(z)$ in Eq.(\ref{int-dilaton}) and $\Phi(z)=\pm z^2$ ,  as examples to show it in an explicit way. Inserting the dilaton profiles and taking $m=0, v_3=0, v_4=8$, one can get the temperature dependent chiral condensate from Eq.(\ref{eom-chi2-s0}). The results are shown in Fig.\ref{check-order}(a). From the figure,  chiral symmetry breaking and restoration are realized in cases with dilaton in Eq.(\ref{int-dilaton}) and $\Phi=-z^2$, while no symmetry breaking appears in case with $\Phi=z^2$. In Fig.\ref{check-order}(b), we calculate $\dot{\chi}_0$ from Eq.(\ref{eq-chi0-se2}) with boundary condition $c_0=1$. From this figure, we could see that in the results of both $\Phi(z)=- z^2$ and
dilaton in Eq.(\ref{int-dilaton}) there is an intersect with $\dot{\chi}_0=0$, while nothing in the results of $\Phi(z)=z^2$. The temperature of the intersects are $T=0.1515...\rm{GeV}$ and $T=0.198...\rm{GeV}$ for dilaton in Eq.(\ref{int-dilaton}) and $\Phi(z)=- z^2$ respectively. Comparing to Fig.\ref{check-order}(a), we could see that they are exactly the same as the second order phase transition temperature in the corresponding cases.  At large $T$ region, $\dot{\chi}_0$ of all the three cases  will approach a limit value  $-\frac{\Gamma^2(\frac{1}{4})}{2\sqrt{\pi}(\psi^{(0)}(\frac{1}{4})-\psi^{(0)}(\frac{3}{4}))}\approx1.18$, which could be extracted by solving the equations setting $\phi(s)\equiv0$. Here, $\psi^{(n)}(z)$ is the $n^{th}$ derivative of the digamma function $\psi(z)\equiv\frac{\Gamma^{'}(z)}{\Gamma(z)}$.

Therefore, by solving the simple linear Eq.(\ref{eq-chi0-se2}), one can check whether the dilaton profile could give a well description of chiral phase transition in chiral limit. The transition temperature could be extracted from Eq.(\ref{eq-chi0-se2}) as well, which would be much easier than from the full solution.  Since Eq.(\ref{eq-chi0-se2}) depends only on the mass term of the scalar interaction, in two-flavor case the second order transition temperature  depends only on dilaton configuration if the metric is fixed to AdS-SW black hole solution. The scalar potential will not affect the location of $\sigma=0$, which is consistent with our previous numerical study in \cite{Chelabi:2015gpc}. This conclusion could be considered as a first test on the dilaton profile.

Up to the leading expansion, one does not have the information about the value of $\epsilon_1$ and $\kappa_1$. So we can turn to higher order expansion. We find that $\epsilon_1=\frac{1}{2}$ and $\kappa_1=\frac{1}{3}$ when taking dilaton in Eq.(\ref{int-dilaton}) and scalar interaction $v_3=0, v_4=8$(For the details, please refer to Appendix.\ref{ana-two}).  Given that $\chi\simeq \frac{\sigma}{\zeta} z^3+o(z^3)$ in chiral limit, one can easy get $\sigma\propto \delta T^{\epsilon_1}$ and $\sigma\propto\delta m^{\kappa_1}$ from the near critical point expansion. So one has $\beta=\epsilon_1=\frac{1}{2},\delta=\kappa_1=\frac{1}{3}$ in the given two-flavor model.  So, the exact value of the critical exponents $\beta$ and $\delta$ do equal to the 3D mean field result $\beta=\frac{1}{2}$ and $\delta=3$. Moreover, we find that for dilaton profile independent on temperature and for scalar interaction $v_3=0, v_4=8$, the critical exponents are always $\beta=\frac{1}{2},\delta=3$, independent on dilaton configuration. It seems that the current model can only produce a mean field result. Here, different from \cite{DeWolfe:2010he}, we are considering a more realistic case with finite $N_f, N_c$ and the suppression of large $N$ effect should not be too large. There should be other reasons that suppress the contributions of the long wave length fluctuations.

\subsection{Go beyond mean field approximation}

\begin{figure}[h]
\begin{center}
\epsfxsize=6.5 cm \epsfysize=6.5 cm \epsfbox{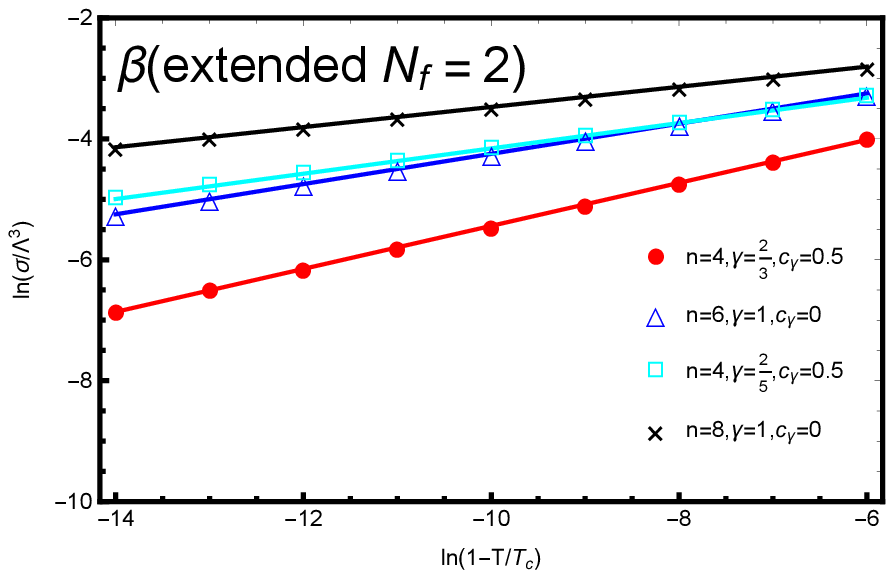}
\hspace*{0.1cm} \epsfxsize=6.5 cm \epsfysize=6.5 cm
\epsfbox{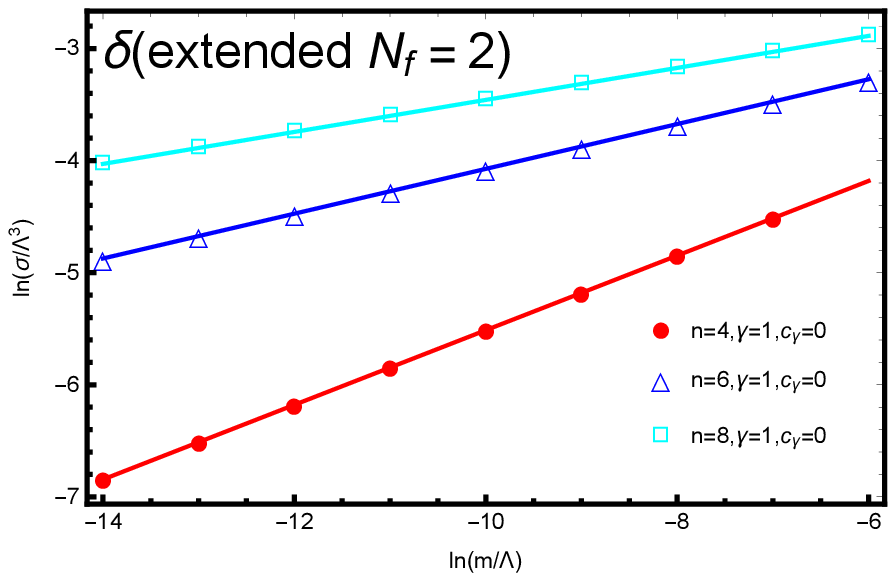} \vskip -0.05cm \hskip 0.15 cm
\textbf{( a ) } \hskip 6.5 cm \textbf{( b )} \\
\end{center}
\caption{Temperature and quark mass dependence of chiral condensate near the critical point $T_{c,2}=0.1515...{\rm MeV}, m_{c,2}=0, \sigma_{c,2}=0$ with the extended dilaton profile in Eq.(\ref{dilaton-critical}) and higher power of scalar potential $\chi^n$. Panel.(a) gives $\ln(\frac{\sigma}{\Lambda})$ as a function of $\ln(1-\frac{T}{T_c})$, with fixed quark mass $m=m_{c,2}=0$. $\Lambda=1\rm{GeV}$ is a parameter introduced to make the quantity inside the logarithmic function dimensionless. The red circle, blue triangle, cyan rectangle and black cross dots are model calculation for $n=4,\gamma=\frac{2}{3}, c_\gamma=0.5$, $n=6,\gamma=1, c_\gamma=0$,$n=4, \gamma=\frac{2}{5}, c_\gamma=0.5$, and $n=4,\gamma=, c_\gamma=0$ respectively.  The solid straight lines are the corresponding linear fitting of the dots, which has the form $y=-1.882+0.356 x$, $y=-1.749+0.250 x$, $y=-2.063+0.209 x$ and $y=-1.806+0.167 x$ respecitvely.  Panel.(b) gives $\ln(\frac{\sigma}{\Lambda})$ as a function of $\ln(\frac{m}{\Lambda})$.   The red circle, blue triangle and cyan rectangle are model calculation for $n=4,\gamma=1, c_\gamma=0$, $n=6,\gamma=1, c_\gamma=0$ and $n=8, \gamma=1, c_\gamma=0$ respectively.  The solid straight lines are the corresponding linear fitting of the dots, which has the form $y=-2.184+0.333 x$, $y=-2.075+0.200 x$ and $y=-2.030+0.143 x$ respecitvely. }
\label{ces-extend}
\end{figure}

Going back to the current model, if we consider the dilaton field representing dynamics from gluon sectors, it is quite natural that it has its dynamics also.  In a full back-reaction framework, one can imagine that the dilaton field should present some kinds of critical scaling behavior, which is neglected in the current model. Furthermore, in the current model, we only consider a simple version of scalar interaction with only the quartic potential, which could be extend to higher order interactions . In order to taking this effect into account, instead of studying the full back-reaction, we take a simple dilaton field
\begin{eqnarray}\label{dilaton-critical}
\phi_T(s)=\Phi(\frac{s}{\pi T})(1+c_\gamma(1-\frac{T}{T_c})^{\gamma})
\end{eqnarray}
to mimic the critical scaling behavior of dilaton field in the full back-reaction scenario. Here $\Phi(z)$ is just the dilaton in Eq.(\ref{int-dilaton},\ref{muvalues}), and $T_c$ is the corresponding transition temperature in chiral limit $T_{c,2}=0.1515...\rm{GeV}$. The scaling exponent $\gamma$ and the coefficient $c_\gamma$ are considered as free parameters. We also consider a more general scalar interaction with n-th power of $\chi$. Then Eq.(\ref{eom-chi2-s0}) becomes
\begin{eqnarray}
\ddot{\chi}-(\frac{3}{s}+\frac{4s^3}{1-s^4}+\dot{\phi}_{T})\dot{\chi}- \frac{1}{s^2(1-s^4)}(3-n v_n\chi^{n-1})=0.\label{eom-chi2-sn}
\end{eqnarray}
Inserting the dilaton field Eq.(\ref{dilaton-critical}) into the above equation and doing the same expansion, one reaches(For the details, please refer to Appendix.\ref{ana-two})
\begin{eqnarray}
\beta=\frac{\gamma}{n-2}, \delta=n-1.
\end{eqnarray}
From the analysis in Appendix.\ref{ana-two}, we find that the coefficient depends only on $\gamma,n$. The coefficient of $v_n$ will not affect the critical exponents in two-flavor case. If one takes $\gamma=1, n=4$, then the results $\beta=\frac{1}{2}, \delta=3$ from the model in the above section is reproduced. In Fig.\ref{ces-extend},  we also test this formula with several groups of $\gamma,c_\gamma, n$ numerically. As mentioned above, $v_n$ would not affect $\beta, \delta$, in all these numerical tests, we take $v_n=1$ for different $n$. From the slope of the linear fitting, we get $\beta=0.356,0.250,0.209, 0.167$ for $n=4,\gamma=\frac{2}{3},\frac{\gamma}{n-2}=\frac{1}{3}\approx0.333$, $n=6,\gamma=1,\frac{\gamma}{n-2}=\frac{1}{4}\approx0.250$, $n=4,\gamma=\frac{2}{5},\frac{\gamma}{n-2}=\frac{1}{5}\approx0.200$, $n=8,\gamma=1,\frac{\gamma}{n-2}=\frac{1}{6}\approx0.166$ respectively.  As for $\delta$, the numerical results are $\delta=3.003,5.000, 6.993$ for $n=4,\gamma=1,n-1=3$, $n=6,\gamma=1,n-1=5$, $n=8,\gamma=1,n-1=7$ respectively. We could see that both the numerical results of $\delta$ agree very well with the analytical result $\delta=n-1$. Up to maximal numerical errors $7\%$, the numerical results of $\beta$ agree with the analytical result $\beta=\frac{\gamma}{n-2}$. The main numerical errors might come from the deviation of the numerical $T_c$ from the exact transition temperature.

As a short summary, though the results in temperature independent dilaton model are still at mean field level, we show the possibility to go beyond mean field approximation using an extended dilaton field with critical scaling and higher power scalar potential. In such an simple extension, one can derive the exact critical exponents as $\beta=\frac{\gamma}{n-2}, \delta=n-1$\footnote{We emphasize that this results are valid only for our simplest settings. In a full back-reaction model, the background metric might be changed and it could also have some kinds of critical scaling in the metric. Here, we just want to use the simple toy model to show the possibility to go beyond mean field approximation.  }. If one build a model with full back-reaction, it is possible to go beyond the mean field approximation. However, this is out of the scope of this work, and we will leave it to the future.

\section{Critical exponents with three degenerate light quarks}
\label{sec-threef}

In this section, we will continue to analyze the case with three degenerate light quarks. So we set $v_3=-3, v_4=8$ as in \cite{Chelabi:2015cwn,Chelabi:2015gpc}. Under the same coordinate transformation in the last section, the equation of motion for the scalar field $\chi$ becomes
\begin{eqnarray}
\ddot{\chi}(s)-(\frac{3}{s}+\frac{4s^3}{1-s^4}+\dot{\phi}_T(s))\dot{\chi}(s)- \frac{1}{s^2(1-s^4)}(3-3v_3\chi^2-4v_4\chi^3)=0.\label{eom-chi3-s}
\end{eqnarray}
Here the dilaton profile would be simply kept as in Eq.(\ref{int-dilaton}). Imposing the boundary condition described in Sec.\ref{sec-soft}, one can solve the temperature and quark mass dependence of chiral condensate. The results are shown in Fig.\ref{sigma-T-degenerate}(b).

\begin{figure}[h]
\begin{center}
\epsfxsize=6.5 cm \epsfysize=6.5 cm \epsfbox{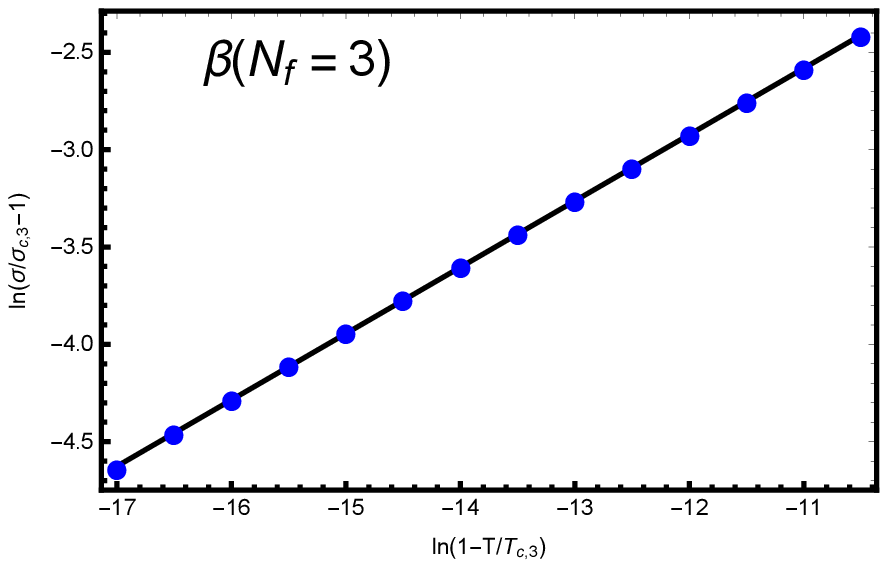}
\hspace*{0.1cm} \epsfxsize=6.5 cm \epsfysize=6.5 cm
\epsfbox{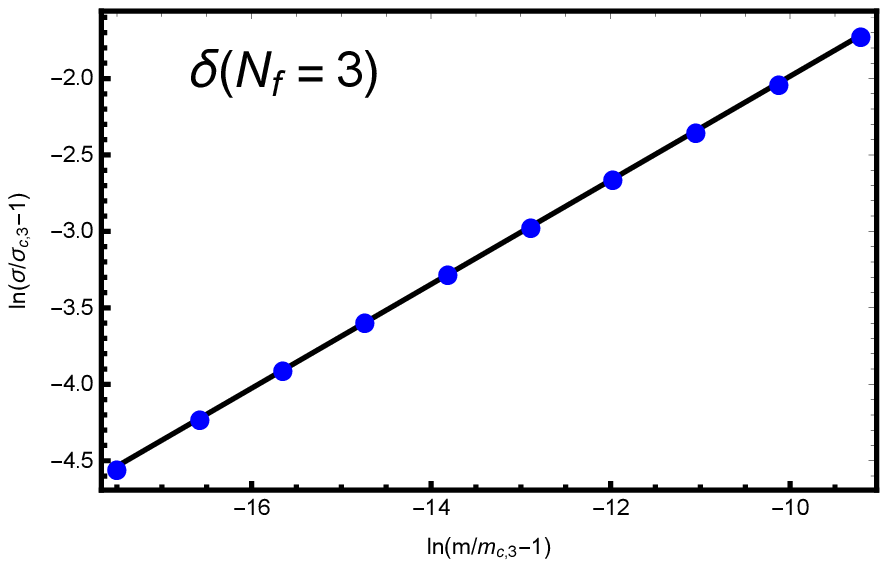} \vskip -0.05cm \hskip 0.15 cm
\textbf{( a ) } \hskip 6.5 cm \textbf{( b )} \\
\end{center}
\caption{Temperature and quark mass dependence of chiral condensate near the critical point $T_{c,3}=0.1888{\rm GeV}, m_{c,3}=0.036871...\rm{GeV}, \sigma_{c,3}=0.027\rm{GeV}^3$ with three degenerate quarks. The dilaton filed is kept as that in Eq.(\ref{int-dilaton}) and the scalar interaction takes the configuration $v_3=-3, v_4=8$. Panel.(a) gives $\ln(\frac{\sigma-\sigma_{c,3}}{\sigma_{c,3}})$ as a function of $\ln(1-\frac{T}{T_{c,3}})$, with fixed quark mass $m=m_{c,3}=0.036871...\rm{GeV}$. The blue dots are model calculation and the black solid straight line are linear fitting of the dots, which has the form $y=1.165+0.340 x$.  Panel.(b) gives $\ln(\frac{\sigma-\sigma_{c,3}}{\sigma_{c,3}})$ as a function of $\ln(\frac{m-m_{c,3}}{m_{c,3}})$.  The blue dots are model calculation and the black solid straight line are linear fitting of the dots, which has the form $y=1.417+0.340 x$. }
\label{ces-threef}
\end{figure}

From Fig.\ref{sigma-T-degenerate}(b), we could see that at small quark mass region $m<m_{c,3}=0.036871...\rm{MeV}$ the phase transition is a first order one, while it turns to be crossover transition only at quark masses larger than $m_{c,3}=0.036871\rm{GeV}$. The critical point of the second order transition locates at $T_{c,3}=0.1888{\rm GeV}, m_{c,3}=0.036871...\rm{GeV}, \sigma_{c,3}=0.027\rm{GeV}^3$ Here, due to the mass effect, chiral condensate is no longer zero in the critical point. In principle, one could add certain counter-terms and get rid of the contribution from finite quark mass. After removing the finite quark mass effect, the critical value of $\sigma$ could be zero. However, since our main goal in this paper is to understand the critical behavior, which should not be affected by the counter-terms, we would not spare efforts on this direction. Instead, we define the critical point where $\frac{d\sigma}{dT}$ diverges and calculate critical exponents from the near critical point scaling behavior.  According to the definition of critical exponent $\beta$  in Eq.(\ref{eqcritical}), we  tune the temperature $T$ slightly around $T_{c,3}=0.1888...{\rm GeV}$ and solve chiral condensate with  fixing quark mass $m_{c,3}=0.36871...\rm{GeV}$. Because the critical behavior of chiral condensate near $T_{c,3}$ obeys certain critical scaling $\sigma-\sigma_{c,3}\propto (T_{c,3}-T)^\beta$. Then one expects that $\ln(\frac{\sigma-\sigma_{c,3}}{\sigma_{c,3}})$ depends linearly on $\ln(t)\equiv\ln(1-\frac{T}{T_{c,3}})$. Here, given that $\sigma_{c,3},T_{c,3}$ are finite, we use them to rescale condensate and temperature deviation to be dimensionless. The slope of the linear line is just the critical exponent $\beta$. Then, in Fig.\ref{ces-threef}(a), we plot $\ln(\frac{\sigma-\sigma_{c,3}}{\sigma_{c,3}})$ as a function of $\ln(t)$ from $\ln(t)=-17$ to $\ln(t)=-10$ . From the plot, it is quite obvious that the model calculated data points(the blue dots ) lie in a straight line, indicating the critical power scaling of chiral condensate. From the best linear fitting, we get the straight line in the form $1.165+0.340x$, which shows $\beta=0.340$.

Similarly, the critical exponent $\delta$ could be extracted from the definition Eq.(\ref{eqcritical}). We tune quark mass $m$ slightly larger than the critical value $m_{c,3}$ and solve $\sigma$ with fixing temperature $T=T_{c,3}$. It is expected that $\sigma-\sigma_{c,3}\propto (m-m_{c,3})^{1/\delta}$, so it is expected that $\ln(\sigma-\sigma_{c,3})$ depends on $\ln(m-m_{c,3})$ linearly. Then, in Fig.\ref{ces-threef}(b), we plot $\ln(\frac{\sigma-\sigma_{c,3}}{\sigma_{c,3}})$ as a function of $\ln(m-m_{c,3})$ from $\ln(m-m_{c,3})=-17$ to $\ln(m-m_{c,3})=-10$ . From the plot, it is quite obvious that the model calculated data points(the blue dots) lie in a straight line, indicating the critical power scaling of chiral condensate with respective to quark mass. From the best linear fitting, we get the straight line in the form $1.417+0.340x$, which shows $\frac{1}{\delta}=0.340$. Therefore, we get $\delta=\frac{1}{0.340}\approx 2.941$.

From the above numerical analysis, we get $\beta=0.340, \delta=2.941$.  As discussed in last section, there could be errors due to the errors in extracting the critical temperature numerically. Thus, we also present an analytical study here. Different from two-flavor case, both the critical values of quark mass and chiral condensate are non-zero. Thus, the critical solution of $\chi$ is non-zero. Therefore, we have to expand the near critical solution as
\begin{eqnarray}
\chi=\chi_{m_c,0}+\chi_{m_c,1}\delta T^{\epsilon_1}+\chi_{m_c,2} \delta T^{\epsilon_2}+...
\end{eqnarray}
under small $T$ deviation $\delta T$ together with fixing $m=m_{c,2}=0$, and
\begin{eqnarray}
\chi=\chi_{m_c,0}+\chi_{T_c,1} \delta m^{\kappa_1}+\chi_{T_c,2} \delta m^{\kappa_2}+...,
\end{eqnarray}
with $\chi_{m_c,0}$ representing the solution at the critical point. After the technical analysis given in Appendix.\ref{ana-three}, we could extract the exact values of the critical exponents, which gives $\beta=\frac{1}{3}\approx 0.333, \delta=3$. It shows that $\beta$ is different from that in two-flavor case while $\delta$ is kept the same.  Furthermore, we note that the value of $\beta$ is quite close to the value $\beta=0.327$ in $Z(2)$ class, while the value of $\delta$ is still far from $\delta=4.789$ in $Z(2)$ class. In some sense, this results might shows that the contribution from the gluon sectors is not as important as in two-flavor case. However, this still requires more test in a full back-reaction model.

\section{Critical exponents in $N_f=2+1$ soft-wall AdS/QCD}
\label{sec-twoplusonef}

In the previous sections, we have studied critical scaling in cases with two and three degenerate flavors. Though in a model without full back-reaction it is still unable to give the exact value of critical exponents as in 4D theory, the results obtained does show that the critical line in two-flavor region and in three-flavor region belong to different universality classes. Thus, in this section, we would like to investigate critical scaling behavior in $N_f=2+1$ case,  the mass plane phase diagram of which has been studied in \cite{Li:2016smq,Bartz:2017jku,Fang:2018vkp}. Here, we will focus on extracting the critical exponents. Before that, for the compactness of this paper, we will give a brief introduction on the results of mass plane phase diagram from soft-wall AdS/QCD.

\begin{figure}[h]
\begin{center}
\epsfxsize=7.5 cm \epsfysize=7.5 cm \epsfbox{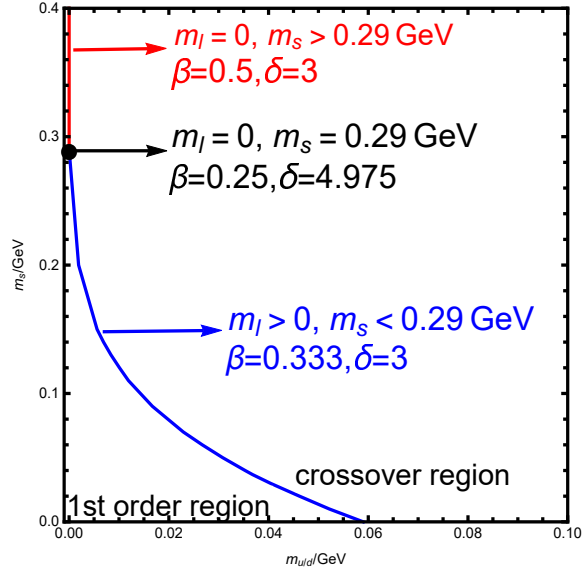} \
\end{center}
\caption{The mass plane($m_{u/d}-m_s$) phase diagram. The lower left corner is first order transition region. The upper right corner is crossover region. The critical line are divided by a tri-critical point $m_{l,tri}=0, m_{s,tri}=0.290...\rm{GeV}$. The blue segment of the critical line has the critical exponents $\beta=\frac{1}{3}\approx0.333,\delta=3$, while the red segment takes  the value $\beta=\frac{1}{2}=0.5,\delta=3$. At the tri-critical point, $\beta=0.25,\delta=4.975$.  } \label{columbia-plot-cal}
\end{figure}

\begin{figure}[h]
\begin{center}
\epsfxsize=6.5 cm \epsfysize=6.5 cm \epsfbox{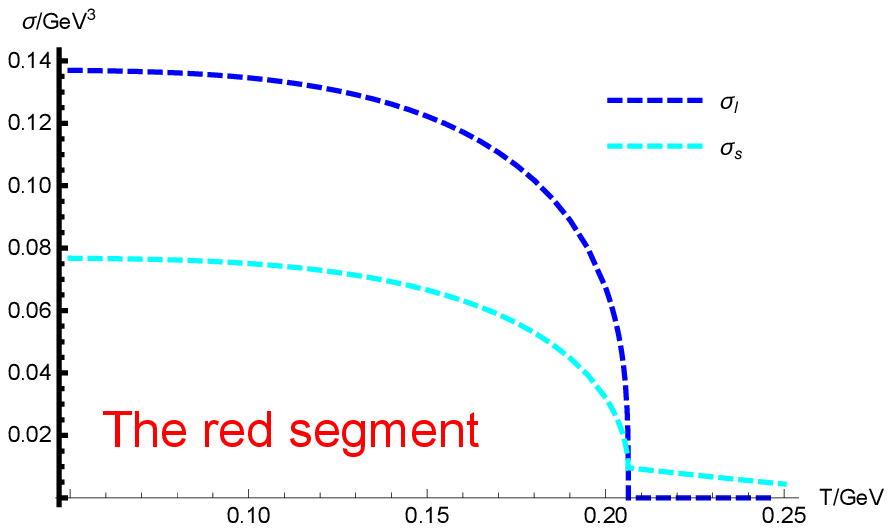}
\hspace*{0.1cm} \epsfxsize=6.5 cm \epsfysize=6.5 cm
\epsfbox{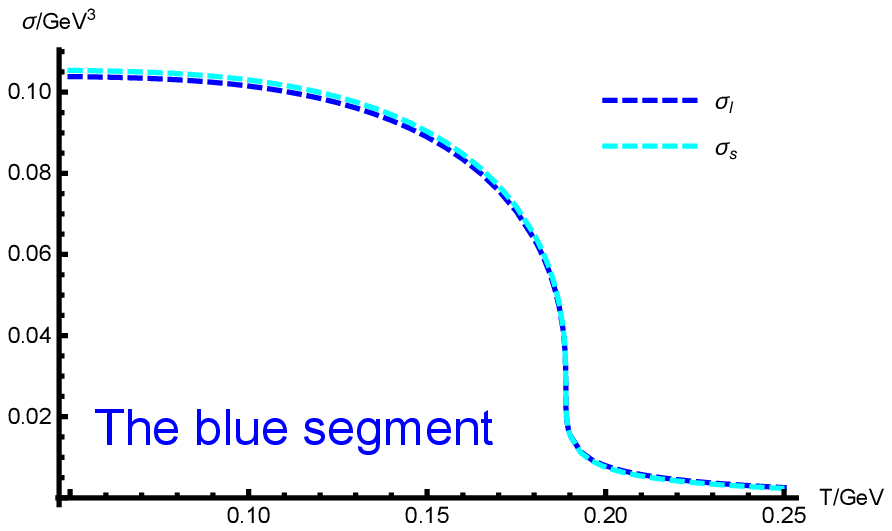} \vskip -0.05cm \hskip 0.15 cm
\textbf{( a ) } \hskip 6.5 cm \textbf{( b )} \\
\end{center}
\caption{Temperature and quark mass dependence of chiral condensate at the red and blue segments of critical line. Panel.(a) gives the result for  $ m_{l,c1}=0, m_{s,c1}=0.300...\rm{GeV}$. The blue and cyan dashed lines represent results of $\sigma_l$ and $\sigma_s$ respectively, which show a second order phase transition. The critical values of  $T, \sigma_l, \sigma_s$ at the phase transition point are $T=0.2063571...{\rm GeV},\sigma_l=0, \sigma_s=0.010\rm{GeV}^3$ . Panel.(b) gives the result for $m_l=0.040\rm{GeV}, m_s=0.030792...\rm{GeV}$. The blue and cyan dashed lines represent results of $\sigma_l$ and $\sigma_s$ respectively, which also show a second order phase transition. The critical values of  $T, \sigma_l, \sigma_s$ at the phase transition point are $T=0.18883...{\rm GeV},\sigma_l=0.0271\rm{GeV}^3, \sigma_s=0.0274\rm{GeV}^3$ .}
\label{sigma-2+1}
\end{figure}

As discussed in Sec.\ref{sec-soft}, when $m_u=m_d\neq m_s$, the expectation value of $X$ field is of the form $X=\text{Diag}\{\chi_l,\chi_l,\chi_s\}$. The equation of motion are given in Eq.(\ref{eom-chil},\ref{eom-chis}). The boundary conditions for $\chi_l,\chi_s$ are given in Eq.(\ref{chiluv},\ref{chisuv},\ref{chiir}). Based on this boundary conditions, one can solve Eq.(\ref{eom-chil},\ref{eom-chis}) with different quark masses and temperature. Taking the dilaton profile in Eq.(\ref{int-dilaton},\ref{muvalues}) and $v_3=-3, v_4=8$, we solved chiral condensate as functions of quark masses and temperature in \cite{Li:2016smq}. We found that below certain critical quark masses, the phase transition is first order, while above the critical value it turns to crossover. All the critical points form a critical line. The results are shown in Fig.\ref{columbia-plot-cal}. In \cite{Li:2016smq}, we only solved the system up to  $m_s=0.2\rm{GeV}$, and we got the blue segment in Fig.\ref{columbia-plot-cal} only. Later, the study in \cite{Bartz:2017jku} solved the results up to a much larger $m_s$, and they found that there is a tri-critical point, above which $\chi_l, \chi_s$ would decoupled at large temperature. This fact was observed in a recent work in \cite{Fang:2018vkp} from a different model. Thus, here, we also extend our calculation to larger $m_s$, and we find that $\chi_l,\chi_s$ does decoupled at high temperature above a tri-critical point at $m_l=0, m_s=0.290\rm{GeV}$. When $m_l=0, m_s>m_{s,tri}$, the solution of $\sigma$ as a function of $T$ would be like the structure in Fig.\ref{sigma-2+1}(a). In the figure, we could see that when $m_l=0, m_s=0.300\rm{GeV}$, above $T=0.2053...\rm{GeV}$, $\sigma_l$ becomes zero and $\chi_l\equiv 0$. It is quite different from the situation at critical line below the tri-critical point. For example, we take another point $m_l=0.010 \rm{GeV}, m_s=0.120105\rm{GeV}<0.290\rm{GeV}$, which locates at the critical line also.  The temperature dependent chiral condensate is shown in Fig.\ref{sigma-2+1}(b). From the figure, we could see that $\sigma_l, \sigma_s$ are strongly coupled with each other even after phase transition. So, we could see that qualitatively the blue segment and the red segment in the critical line are quite different. At the red segment, above the the transition temperature both $\sigma_l$ and $\sigma_s$ vanishes.

\begin{figure}[h]
\begin{center}
\epsfxsize=6.5 cm \epsfysize=6.5 cm \epsfbox{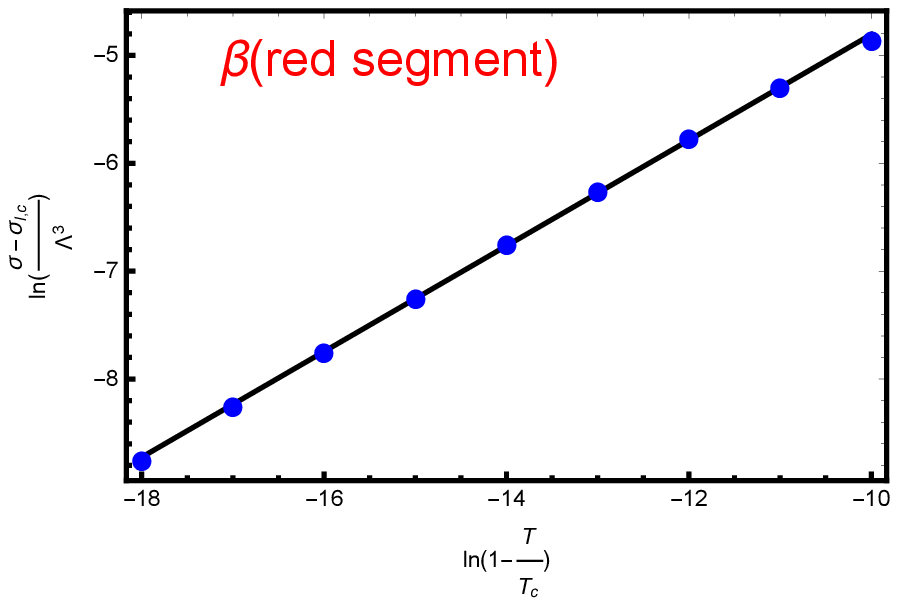}
\hspace*{0.1cm} \epsfxsize=6.5 cm \epsfysize=6.5 cm
\epsfbox{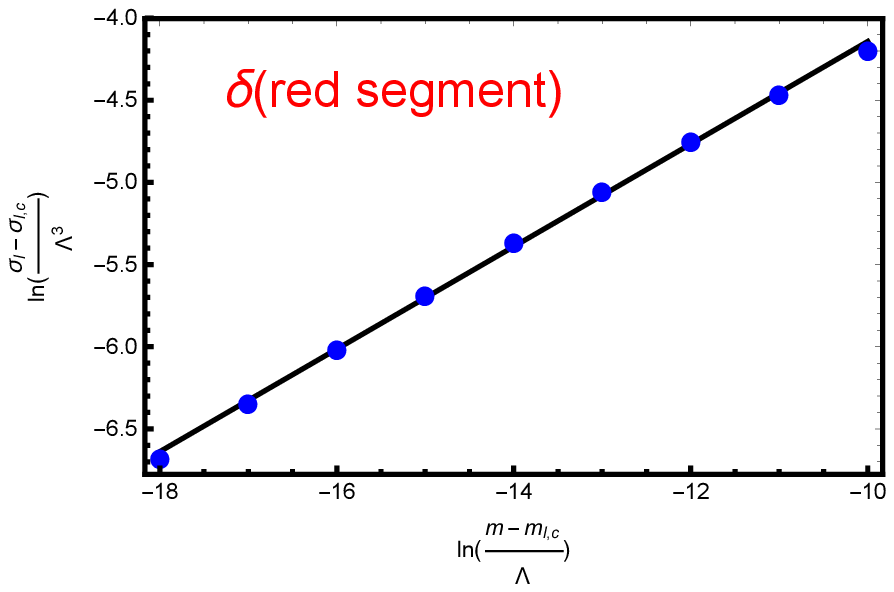} \vskip -0.05cm \hskip 0.15 cm
\textbf{( a ) } \hskip 6.5 cm \textbf{( b )} \\
\end{center}
\caption{Temperature and quark mass dependence of chiral condensate near the critical point $T_{l,c1}=0.2063571...{\rm GeV}, m_{l,c1}=0, m_{s,c1}=0.300...\rm{GeV}, \sigma_{l,c1}=0, \sigma_{s,c1}=0.010\rm{GeV}^3$, which locates at the red segment of the chiral critical line. Panel.(a) gives $\ln(\frac{\sigma_l-\sigma_{l,c1}}{\Lambda^3})=\ln(\frac{\sigma_l}{\Lambda})$ as a function of $\ln(t)\equiv\ln(\frac{T_{l,c1}-T}{T_{l,c1}})$, with fixed quark mass $m_l=m_{l,c1}=0, m_s=m_{s,c1}=0.300...\rm{GeV}$. The blue dots are model calculation and the black solid straight line are linear fitting of the dots, which has the form $y=0.093+0.490 x$.  Panel.(b) gives $\ln(\frac{\sigma_l-\sigma_{l,c1}}{\Lambda^3})=\ln(\frac{\sigma_l}{\Lambda})$ as a function of $\ln(\frac{m_l-m_{l,c1}}{\Lambda})=\ln(\frac{m_l}{\Lambda})$.  The blue dots are model calculation and the black solid straight line are linear fitting of the dots, which has the form $y=-1.022+0.312 x$.}
\label{ces-2+1-1}
\end{figure}

To see the quantitative difference of the two segments, we would try to investigate the critical exponents here. Firstly, we take a point $m_{l,c1}=0, m_{s,c1}=0.300$, which locates at the red segment of the chiral critical line. The critical values at the second order phase transition point are $T_{l,c1}=0.2063571...{\rm GeV},\sigma_{l,c1}=0, \sigma_{s,c1}=0.010\rm{GeV}^3$. Like in previous sections, we plot $\ln(\frac{\sigma_l-\sigma_{l,c1}}{\Lambda^3})=\ln(\frac{\sigma_l}{\Lambda^3})$ as a function of $\ln(t)\equiv\ln(\frac{T_{l,c1}-T}{T_{l,c1}})$ and $\ln(\frac{m_l-m_{l,c1}}{\Lambda})=\ln(\frac{m_l}{\Lambda})$ in Fig.\ref{ces-2+1-1}(a) and (b) respectively. Here $\Lambda=1\rm{GeV}$ is a parameter introduced to make the quantities inside logarithmic function dimensionless. Fig.\ref{ces-2+1-1}(a) and (b), we could see that near the second order phase transition point $\sigma$ have good scaling behavior. The data points $(\ln(t), \ln(\frac{\sigma_l-\sigma_{l,c1}}{\Lambda^3}))$ and $(\ln(\frac{m_l-m_{l,c1}}{\Lambda}),\ln(\frac{\sigma_l-\sigma_{l,c1}}{\Lambda^3}))$ lies in straight lines of forms $y=0.093+0.490 x$ and $y=-1.022+0.312 x$ respectively.  This gives $\beta=0.492,\delta=\frac{1}{0.312}\approx3.205$. In fact, since in the red segment of critical line, near the transition temperature, $\chi_l,\chi_s$ decouple with each other. So it is easy to understand that the exact value of critical exponent should be the same as in two-flavor case. That is $\beta=\frac{1}{2},\delta=3$. The numerical calculation has checked this exact value up to $7\%$ numerical errors.

\begin{figure}[h]
\begin{center}
\epsfxsize=6.5 cm \epsfysize=6.5 cm \epsfbox{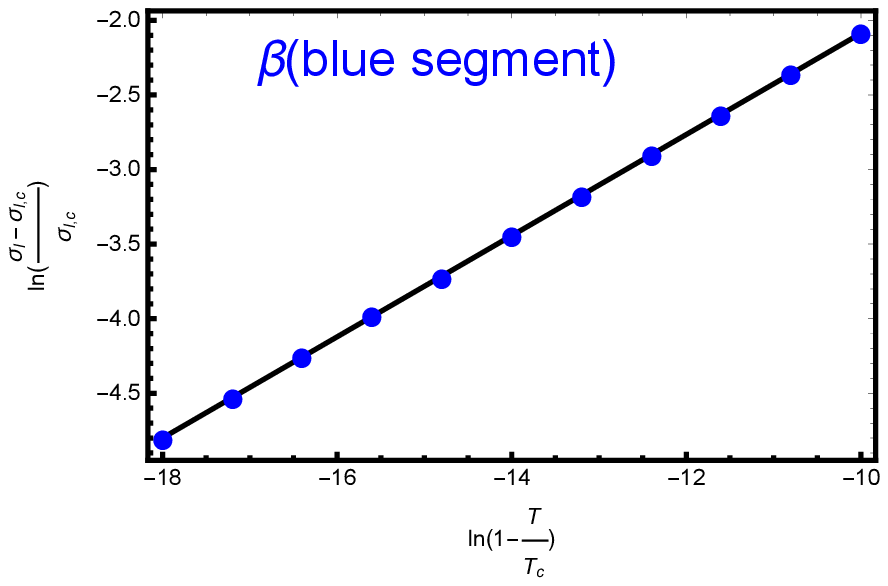}
\hspace*{0.1cm} \epsfxsize=6.5 cm \epsfysize=6.5 cm
\epsfbox{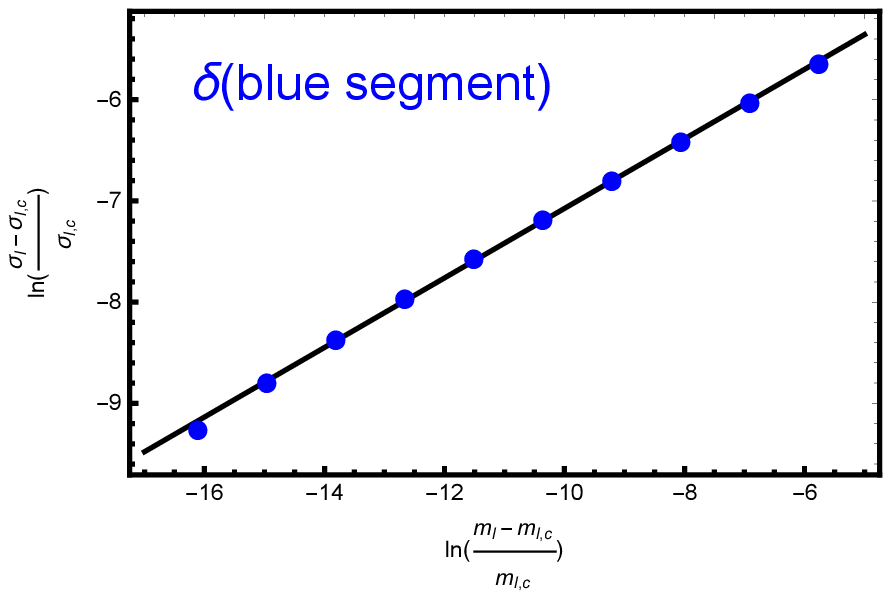} \vskip -0.05cm \hskip 0.15 cm
\textbf{( a ) } \hskip 6.5 cm \textbf{( b )} \\
\end{center}
\caption{Temperature and quark mass dependence of chiral condensate near the critical point $T_{l,c2}=0.19123136679...{\rm GeV}, m_{l,c2}=0.010\rm{GeV}, m_{s,c2}=0.120105...\rm{GeV}, \sigma_{l,c2}=0.0271...\rm{GeV}^3, \sigma_{s,c2}=0.0232...\rm{GeV}^3$, which locates at the blue segment of the chiral critical line.. Panel.(a) gives $\ln(\frac{\sigma_l-\sigma_{l,c2}}{\sigma_{l,c2}})$ as a function of $\ln(1-\frac{T}{T_{l,c2}})$, with fixed quark mass $m_l=m_{l,c2}=0.010\rm{GeV}, m_s=m_{s,c2}=0.120105...\rm{GeV}$. The blue dots are model calculation and the black solid straight line are linear fitting of the dots, which has the form $y=1.304+0.339 x$.  Panel.(b) gives $\ln(\frac{\sigma_l-\sigma_{l,c2}}{\sigma_{l,c2}})$ as a function of $\ln(\frac{m_l-m_{l,c2}}{m_{l,c2}})$.  The blue dots are model calculation and the black solid straight line are linear fitting of the dots, which has the form $y=-3.63952+0.343 x$. }
\label{ces-2+1-2}
\end{figure}

Then, we take another point $m_{l,c2}=0.010\rm{GeV},m_{s,c2}=0.120105...\rm{GeV}$, locating at the blue segment of the critical line. Like in three-flavor case, at high temperature, chiral condensate is not exactly zero due to the explicit symmetry breaking from finite quark masses. Thus, we define the transition temperature where  $\frac{d\sigma}{dT}$ diverges. According to this definition, we determine the phase transition point at $T_{l,c2}=0.19123136679...{\rm GeV},\sigma_{l,c2}=0.0271...\rm{GeV}^3, \sigma_{s,c2}=0.0232...\rm{GeV}^3$. In Fig.\ref{ces-2+1-2}(a) and (b), we plot $\ln(\frac{\sigma_l-\sigma_{l,c2}}{\sigma_{l,c2}})$ as functions of $\ln(1-\frac{T}{T_{l,c2}})$ and $\ln(\frac{m_l-m_{l,c2}}{m_{l,c2}})$ respectively. We could see that all the data points lie in straight lines, which show the good critical scaling behavior. From linear fitting, the straight lines have the form $y=1.304+0.339 x$ and $y=-3.63952+0.343 x$ for $\beta$ and $\delta$ calculation respectively. It gives $\beta=0.339, \gamma=\frac{1}{0.343}\approx2.915$, very close to the exact value of three-flavor. Actually, one can get the exact value $\beta=\frac{1}{3},\delta=3$ analytically, using the same method discussed in Appendix.\ref{ana-three}.  The numerical results and the analytical results match with each other up to $3\%$ numerical errors.

\begin{figure}[h]
\begin{center}
\epsfxsize=6.5 cm \epsfysize=6.5 cm \epsfbox{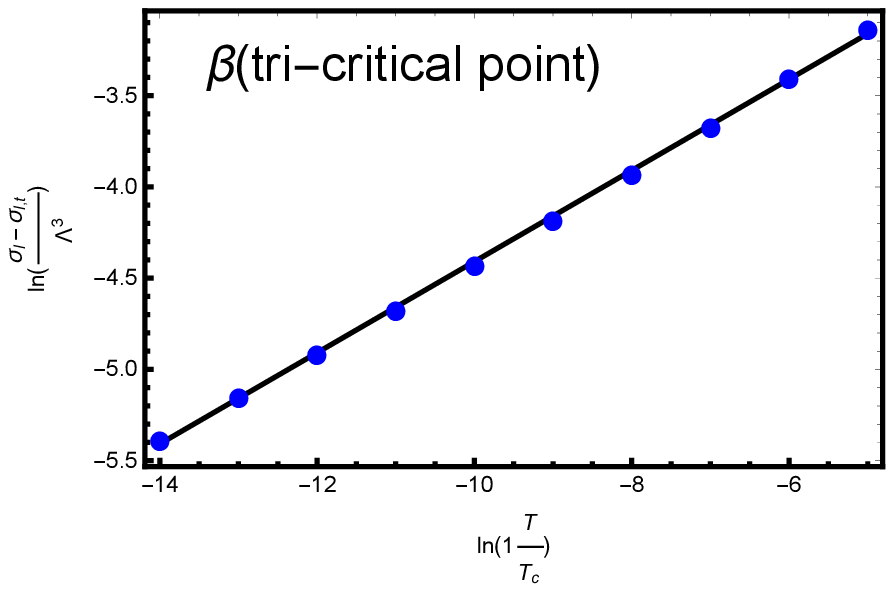}
\hspace*{0.1cm} \epsfxsize=6.5 cm \epsfysize=6.5 cm
\epsfbox{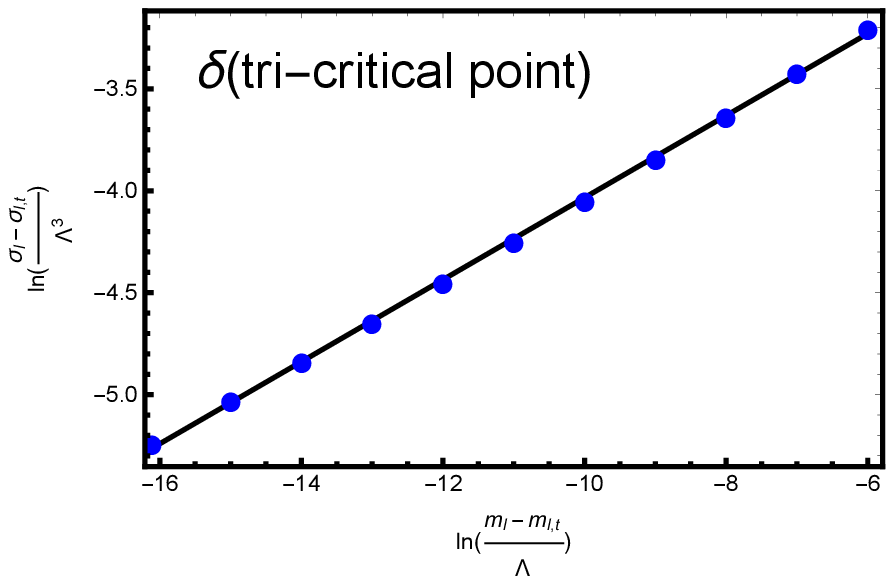} \vskip -0.05cm \hskip 0.15 cm
\textbf{( a ) } \hskip 6.5 cm \textbf{( b )} \\
\end{center}
\caption{Temperature and quark mass dependence of chiral condensate near the tri-critical point $T_{l,t}=0.2053135...{\rm GeV}, m_{l,t}=0, m_{s,t}=0.290...\rm{GeV}, \sigma_{l,c}=0, \sigma_{s,t}=0.010\rm{GeV}^3$. Panel.(a) gives $\ln(\frac{\sigma_l-\sigma_{l,t}}{\Lambda^3})$ as a function of $\ln(1-\frac{T}{T_c})$, with fixed quark mass $m_l=m_{l,t}=0, m_s=m_{s,t}=0.290...\rm{GeV}$. $\Lambda$ is taken as $1\rm{GeV}$. The blue dots are model calculation and the black solid straight line are linear fitting of the dots, which has the form $y=-1.912+0.250 x$.  Panel.(b) gives $\ln(\frac{\sigma_l-\sigma_{l,t}}{\Lambda^3})$ as a function of $\ln(\frac{m_l-m_{l,t}}{\Lambda})$.  The blue dots are model calculation and the black solid straight line are linear fitting of the dots, which has the form $y=-2.022+0.201 x$. }
\label{ces-2+1-c}
\end{figure}

Finally, we turn to the tri-critical point with $m_{l,t}=0, m_{s,t}=0.290...\rm{GeV}$. Solving the equation of motion, we find that the second order phase transition point locate at $T_{l,t}=0.2053135...{\rm GeV}, \sigma_{l,c}=0, \sigma_{s,t}=0.010\rm{GeV}^3$. In Fig.\ref{ces-2+1-1}(a) and (b),  we plot $\ln(\frac{\sigma_l-\sigma_{l,t}}{\Lambda^3})=\ln(\frac{\sigma_l}{\Lambda^3})$ as a function of $\ln(t)\equiv\ln(\frac{T_{l,t}-T}{T_{l,t}})$ and $\ln(\frac{m_l-m_{l,t}}{\Lambda})$  respectively. Again, we find that all the data points lie in straight lines, verifying the critical scaling behavior. From the linear fitting, we get the straight lines of the forms $y=-1.912+0.250 x$ and $y=-2.022+0.201 x$, for $\beta$ and $\delta$ respectively. Accordingly, we get $\beta=0.250,\delta=\frac{1}{0.201}\approx4.975$. It is interesting to note that $\delta$ at the tri-critical point is quite close to the value $\delta=4.824$ in $O(4)$ class and $\delta=4.789$ in $Z(2)$ class, while $\beta =0.333$ is quite close to $\beta=0.385$ in $O(4)$ class and $\beta=0.327$ in $Z(2)$ class. Probably, the full back-reaction study could have important improvement. We will leave it to the future.

As a short summary, we study the critical scaling behavior of chiral condensate in $N_f=2+1$ model with $m_u=m_d\neq m_s$. We get $\beta=0.5, \delta=3$ in the red segment, $\beta=0.333,\delta=3.0$ in the blue segment, and $\beta=0.250,\delta=4.975$ in the tri-critical point. We also label this results in Fig.\ref{columbia-plot-cal}.  This results confirm that the red segment, blue segment and tri-critical point in the chiral critical line belong to different universality classes.

\section{Conclusion and discussion}
\label{sec-sum}

\begin{table}
\begin{center}
\begin{tabular}{cccccccc}
\hline\hline
          ~                        &  $N_f=2$                & extended $N_f=2$                     & $N_f=3$             \\
\hline

        $\beta$                    & $\frac{1}{2}$          & $\frac{\gamma}{n-2}$   &   $\frac{1}{3}$                \\

       $\delta$                    & 3                       &$n-1$                  &  $3$        &   \\
\hline\hline
       $N_f=2+1$                  &$m_s>m_{s,t}$        & $m_s=m_{s,t}$    &$m_s<m_{s,t}$\\
\hline
       $\beta$                    &$\frac{1}{2}$             &$0.250$                 &$\frac{1}{3}$\\
       $\delta$                   &3                         &$4.975$                 &3\\
\hline
\end{tabular}
\caption{Summarization of critical exponents in different cases from soft-wall AdS/QCD model. Except for the tri-critical point(only numerical result),  all the results are exact values checked by analytical analysis in the Appendix.}
\label{crit-sum}
\end{center}
\end{table}

In this work, we investigate chiral phase transition in soft wall AdS/QCD models. Based on our study in \cite{Chelabi:2015cwn,Chelabi:2015gpc,Li:2016smq}, we focus on extracting the critical scaling behavior of chiral condensate. Starting from a general $SU(N_f)\times SU(N_f)$ soft wall model, we derive the effective action and equations of motion in different situations, especially for two, three degenerate quarks and $N_f=2+1$ with $m_u=m_d\neq m_s$. Then we solve the critical points and calculate the critical exponents around the critical points. The results are summarized in Table.\ref{crit-sum}.

For $N_f=2$ case, we calculate the critical exponents both numerically and analytically. The exact value of critical exponents are $\beta=\frac{1}{2},\delta=3$ in this case. It equals to the 3D mean field calculation  exactly.  Since in some sense the holographic approach have taken the running with energy scale into account, it is quite unreasonable that there is no correction from the mean field approximation. After careful study, it is found that the full back-reaction might be important to improve the result. Since the current study is in probe limit and the dilaton field is input by hand, there is no temperature dependence in the dilaton. Thus, we have not considered the running effect with temperature correctly. Instead of doing a full back-reaction analysis, we take a simple dilaton configuration $\Phi_1=\Phi(1+c_\gamma(1-\frac{T}{T_c})^\gamma)$ to mimic the possible critical behavior of dilaton field due to the full back-reaction. Furthermore, we extend the quartic interacting term $|X|^4$ to higher power $|X|^n$. Then we get the critical exponents analytically. It gives $\beta=\frac{\gamma}{n-2}, \delta=n-1$, which could be reduced to $\beta=\frac{1}{2},\delta=3$ when taking $\gamma=1,n=4$. Of course, this toy model is still a very rough one, omitting the critical scaling of metric and other terms. But it does show that the full back-reaction could have important improvement on the critical exponents.  As a byproduct, when deriving the expression of critical exponents, we find a simple criteria of dilaton field, on whether it can produce the correct behavior of chiral phase transition in two-flavor case. One can simply solve the linear Eq.(\ref{eq-chi0-se2}) and check whether there is a temperature giving $\dot{\chi}_{m_c,1}(s=0)=0$. Since the linear equation would be much easier to solve numerically, this criteria could provide a simple first test on the dilaton field, as well as a simple calculation on the transition temperature.

For $N_f=3$ case, we also calculate the critical exponents by numerical method and analytical analysis. The exact value of critical exponents are $\beta=\frac{1}{3},\delta=3$ in this case. The different values of critical exponents from $N_f=2$ case shows that they belongs to different universality classes. From the analytical calculation, we find that the critical exponents would not depend on the coefficients of the scalar potential. It is consistent with the fact that the critical exponents are independent on the interacting details.

Then, for $N_f=2+1$, we find that the chiral critical line are divided into two different segments by a tri-critical point at $m_l=0, m_s=0.290\rm{GeV}$. The segment above the tri-critical point belongs to the same universality class as the two-flavor limit with $\beta=\frac{1}{2},\delta=3$, while the segment below the tri-critical point belongs to the same universality classes as $N_f=3$ case with $\beta=\frac{1}{3},\delta=3$. Furthermore, the tri-critical point belongs to another universality class with $\beta=0.25,\delta=4.975$.

The study in this work provides a preliminary analysis on chiral criticality. Qualitatively, the phase diagram agrees with that from 4D analysis. The critical exponents in certain cases is close to the 4D results. However, the model studied here is still in probe limit and in some cases it can not take the running behavior with temperature and other quantities into account. Therefore, it is quite interesting to consider a full back-reaction model based on this probe limit study.  The study here does show that there is possibility to build a correct full back-reaction model based on the soft wall model. We will leave this to the future.

\vskip 0.5cm
{\bf Acknowledgement}
\vskip 0.2cm
We would like to thank the useful discussion with Heng-Tong Ding and Bin Qin. J.C. and D.L. is supported by the NSFC under Grant Nos. 11805084 and 11647141. S.H. is supported from Max-Planck fellowship in Germany and the German-Israeli Foundation for Scientific Research and Development. M.H. is supported by the NSFC under Grant Nos. 11725523, 11735007 and 11261130311(CRC 110 by DFG and NSFC). This work is also partly supported by the PhD Start-up Fund of Natural Science Foundation of Guangdong Province(No.2018030310457).

\appendix

\section{Extracting critical exponents analytically }
\label{appd}
\subsection{Expansion around $\chi=0$}
\label{ana-two}
In this section, we would present a analytical derivation on the critical exponents $\beta,\delta$ in soft-wall model with two degenerate quarks . The start point would be Eq.(\ref{eom-chi2}).  For later convenience,  we firstly make a coordinate transformation $z= z_h s=\frac{s}{\pi T}$ to Eq.(\ref{eom-chi2}), under which the horizon would be set to $s=1$ and the boundary is at $s=0$. Doing this coordinate transformation, Eq.(\ref{eom-chi2}) becomes
\begin{eqnarray}
\ddot{\chi}-(\frac{3}{s}+\frac{4s^3}{1-s^4}+\dot{\phi}_T)\dot{\chi}- \frac{1}{s^2(1-s^4)}(3-4v_4\chi^3)=0.\label{eom-chi2-s}
\end{eqnarray}
Here the 'dot'  stands for derivative with respective to $s$ and $\phi_T(s)\equiv \Phi(\frac{s}{\pi T})$ are dilaton field in the new coordinate $s$. To get a general conclusion, here we consider $\phi_T(s)$ as a general regular function.  Then, as we mentioned in the introduction, in chiral limit of two-flavor system, chiral phase transition would be second order. A successful holographic description should satisfy this condition first. Therefore, at the critical point both $m$ and $\sigma$ should be zero for two-flavor model. Thus the solution of $\chi$ should be zero at the critical temperature $T_c$ also. If $T$ slightly deviates from $T_c$ from the left hand side($T<T_c$), we expect that $\chi$ would deviate from $\chi=0$ slightly, of a certain power of the temperature deviation $\delta T\equiv T_c-T$. So we expand the solution of $\chi$ at $T_c$ as
\begin{eqnarray}
\chi=\chi_{m_c,1}\delta T^{\epsilon_1}+\chi_{m_c,2} \delta T^{\epsilon_2}+....\label{expan-two-T}
\end{eqnarray}
Here $\epsilon_1$ and $\epsilon_2$ are two positive real number and without loss of generality we assume $\epsilon_1<\epsilon_2$. $\chi_{m,1}$ and $\chi_{m,2}$ are two functions of $s$, independent on $T$. The lower index $m_c$ means fixing $m=m_c$.  Inserting Eq.(\ref{expan-two-T}) into Eq.(\ref{eom-chi2-s}) and keeping only the leading expansion on $T_c-T$, we get
\begin{eqnarray}\label{eq-chi0}
\ddot{\chi}_{m_c,1}-(\frac{3}{s}+\frac{4s^3}{1-s^4}+\dot{\phi}_{T_c})\dot{\chi}_{m_c,1}+ \frac{3}{s^2(1-s^4)}\chi_{m_c,1}=0.
\end{eqnarray}
The above equation is a second order linear equation. The near horizon expansion of $\chi_{m_c,1}(s)$ could be easily extracted as
\begin{eqnarray}\label{bd-c0}
\chi_{m_c,1}(s)=c_0(1+\frac{3}{4}(s-1))+c_1 \ln(1-s)(1+\frac{3}{4}(s-1))-c_1((3+\dot{\phi}_{T_c}(1))(1-s)),
\end{eqnarray}
with $c_0, c_1$ the two integral constants. It is obvious that the $c_1$ branch is divergent at the horizon $s=1$. Thus, the regularity of $\chi_{m_c,1}$ requires $c_1=0$. Imposing this condition, we can solve the full solution of $\chi_{m_c,1}(s)$ numerically and extract $\dot{\chi}_{m_c,1}(s=0)$ , which is proportional to the quark mass and equals to zero in chiral limit. Then it is easy to understand that if there exists a second order phase transition point in chiral limit, there should exist a non-trivial solution of Eq.(\ref{eq-chi0}) satisfying both $\dot{\chi}_{m_c,1}(s=0)$ and Eq.(\ref{bd-c0}) with $c_1=0$. Actually, this condition could be used as a first test of the holographic model setting, especially to check whether the correct behavior of chiral symmetry breaking and restoration in chiral limit has been well described or not. In Sec.\ref{sec-twof}, we have taken several examples to show this conclusion in a more explicit way. So we would not repeat it here.

The leading order expansion of Eq.(\ref{eom-chi2-s}) has the information about the second order phase transition point. However, there is no information about the critical exponent $\beta$. In order to get the exact value of $\beta$, we have to go to the next leading order. Before that, we assume a more general form of the dilaton field
\begin{eqnarray}
\phi_T(s)\simeq\phi_{T_c}(s)+g_{T_c}(s)\delta T^{\gamma},
\end{eqnarray}
which can be considered as the near $T_c$ expansion of the dilaton field in a full back-reaction model. The critical scaling of dilaton field is control by $\gamma$.  To reach more general conclusion, we also extend the quartic scalar interaction $v_4\chi^4$ to a general power $v_n \chi^n$ in the discussion here. Given these and assuming $\epsilon_2=\gamma+\epsilon_1=(n-1)\epsilon_1$, i.e. $\epsilon_1=\frac{\gamma}{n-2}$, the leading order expansion are kept as Eq.(\ref{eq-chi0}),  while the next to leading order becomes
\begin{eqnarray}\label{eq-chi1}
\ddot{\chi}_{m_c,2}-(\frac{3}{s}+\frac{4s^3}{1-s^4}+\dot{\phi}_{T_c}))\dot{\chi}_{m_c,2}+ \frac{3}{s^2(1-s^4)}\chi_{m_c,2}=g_{T_c}\dot{\chi}_{m_c,1}+ \frac{nv_n\chi_{m_c,1}^{n-1}}{s^2(1-s^4)}.
\end{eqnarray}
Actually, it is just Eq.(\ref{eq-chi0}) with an additional source $J(s)\equiv g_{T_c}\dot{\chi}_{m_c,1}+ \frac{nv_n\chi_{m_c,1}^{3}}{s^2(1-s^4)}$. The general solutions of $\chi_2$ would have the form
\footnote{Generally, if $y_1(x), y_2(x)$ are two independent solutions of the second order derivative equation $Y^{''}(x)+q(x)Y^{'}(x)+h(x)Y(x)=0$, then one can prove that the general solutions of $Y^{''}(x)+q(x)Y^{'}(x)+h(x)Y(x)=J(x)$ would be $Y(x)=y_1(x)\int_{x}^{x_1}\frac{y_2(x)J(x)}{y_1(x)y_2^{'}(x)-y_1^{'}(x)y_2(x)}+y_2(x)\int_{x_2}^{x}\frac{y_1(x)J(x)}{y_1(x)y_2^{'}(x)-y_1^{'}(x)y_2(x)}$ with $x_1,x_2$ related to the two integral constants, which should be fixed by boundary conditions.}
\begin{eqnarray}\label{chi-1}
\chi_{m_c,2}(s)=&&y_1(s)\int_{s}^{s_1} d\tau \frac{y_2(\tau)\left(g_{T_c}(\tau)\dot{y}_2(\tau)+\frac{nv_n(y_2(\tau))^{n-1}}{\tau^2(1-\tau^4)}\right)}{y_1(\tau)\dot{y}_2(\tau)-\dot{y}_1(\tau)y_2(\tau)}\nonumber\\
&&+y_2(s)\int_{s_2}^{s} d\tau \frac{y_1(\tau)\left(g_{T_c}(\tau)\dot{y}_2(\tau)+\frac{nv_n(y_2(\tau))^{n-1}}{\tau^2(1-\tau^4)}\right)}{y_1(\tau)\dot{y}_2(\tau)-\dot{y}_1(\tau)y_2(\tau)},
\end{eqnarray}
with $y_1(s), y_2(s)$ two independent solutions of Eq.(\ref{eq-chi0}). Since we assume that we are expanding around a critical point with $m_c=0,\sigma_c=0,\chi_c=\equiv0$, Eq.(\ref{eq-chi0}) should have a non-vanishing solution $\chi_1(s)$ with boundary condition $\dot{\chi}_1(0)=0$ and $\chi_1(1)=1$. Therefore, without loss of generality, we could assume $y_2(s)=c \chi_1(s)$.  It is easy to see that the other branch $y_1(s)$ satisfies the boundary condition $y_1(s)= c_1 \log(1-s)+o(1)$ near $s=1$ and $y_1(s)\propto s$ near the boundary $s=0$.  In order to make sure that $\chi_{m_c,2}(s)$ in Eq.(\ref{chi-1}) is regular near $s=1$,  one should fix the integral constant $s_1=1$, which gets rid of the singular branch $y_1(s)$ at the boundary.  Furthermore,  to make sure that we are considering chiral limit(here we need $\dot{\chi}_{m_c,2}(0)=0$), the following condition is required
\begin{eqnarray}\label{nonsingular}
\int_{0}^{1} d\tau \frac{g_{T_c}(\tau)\dot{y}_2(\tau)y_2(\tau)+\frac{nv_n(y_2(\tau))^{n}}{\tau^2(1-\tau^4)}}{y_1(\tau)\dot{y}_2(\tau)-\dot{y}_1(\tau)y_2(\tau)}=0.
\end{eqnarray}
Then, inserting $y_2(s)=c \chi_1(s)$ into Eq.(\ref{nonsingular}), we have
\begin{eqnarray}
\int_{0}^{1} d\tau  \frac{g_{T_c}(\tau)\chi_1(\tau)\dot{\chi}_1(\tau)}{y_1(\tau)\dot{\chi}_1(\tau)-\dot{y}_1(\tau)\chi_1(\tau)}=-c^{n-2}n\lambda\int_{0}^{1} d\tau \frac{(\chi_1(\tau))^{n}}{\tau^2(1-\tau^4)(y_1(\tau)\dot{\chi}_1(\tau)-\dot{y}_1(\tau)\chi_1(\tau))},\nonumber\\
\end{eqnarray}
from which we can have
\begin{eqnarray}
c^{n-2}n\lambda=-\int_{0}^{1} d\tau \frac{(\chi_1(\tau))^{n}}{\tau^2(1-\tau^4)(y_1(\tau)\dot{\chi}_1(\tau)-\dot{y}_1(\tau)\chi_1(\tau))}/\int_{0}^{1} d\tau  \frac{g_{T_c}(\tau)\chi_1(\tau)\dot{\chi}_1(\tau)}{y_1(\tau)\dot{\chi}_1(\tau)-\dot{y}_1(\tau)\chi_1(\tau)}.\nonumber\\
\end{eqnarray}
Since $\chi_0(t)$ and $y_1(t)$ depends only on the linear equation Eq.(\ref{eq-chi0}), it is easy to understand $c^{n-2}n\lambda$ depends only on configuration of $\Phi$. Rewriting in coordinate $z$,  $\chi_1(t)=c t^3=c (\pi T_c)^3 z^3=\frac{\sigma}{\zeta}z^3$ near $t=0$, with $\zeta=\frac{\sqrt{3}}{2\pi}$ for the number of colors $N_c=3$\cite{Cherman:2008eh}. Then the critical behavior of chiral condensate $\sigma(T)\simeq \frac{\sqrt{3}c}{2\pi } (\pi T_c)^3 (T_c-T)^{\frac{\gamma}{n-2}}$, where the critical exponent $\frac{\gamma}{n-2}$ depends not only on $\gamma$ but also on $n$. The slope of the critical scaling equals $\frac{\sqrt{3}c}{2\pi } (\pi T_c)^3=\frac{\sqrt{3}}{2\pi } (\pi T_c)^3 (\frac{C}{n\lambda})^{\frac{1}{n-2}}$, with
\begin{eqnarray}
C\equiv -\int_{0}^{1} d\tau \frac{(\chi_1(\tau))^{n}}{t^2(1-t^4)(y_1(\tau)\dot{\chi}_1(\tau)-\dot{y}_1(\tau)\chi_1(\tau))}/\int_{0}^{1} d\tau  \frac{g_{T_c}(\tau)\chi_1(\tau)\dot{\chi}_1(\tau)}{y_1(\tau)\dot{\chi}_1(\tau)-\dot{y}_1(\tau)\chi_1(\tau)}.
\end{eqnarray}

The above derivation depends on the assumption $\epsilon_2=\gamma+\epsilon_1=(n-1)\epsilon_1$, i.e. $\epsilon_1=\frac{\gamma}{n-2}$. Actually, one can easy understands that this is the only way to get non-trivial branch of $\sigma$ near $T_c$. If $\epsilon_2=\gamma+\epsilon_1<(n-1)\epsilon_1$ or $\epsilon_2=(n-1)\epsilon_1<\gamma+\epsilon_1$, then one of the two terms in Eq.(\ref{nonsingular}) would be absence, which would stop the cancelation of the two terms and Eq.(\ref{nonsingular}) can not be satisfied. Therefore, in fact, $\epsilon_2=\gamma+\epsilon_1=(n-1)\epsilon_1$ is a necessary condition. So,  we have prove that $\epsilon_1=\frac{\gamma}{n-2}$. Considering $\chi_{m_c,1}(s)\propto \sigma$, we could get $\beta=\epsilon_1=\frac{\gamma}{n-2}$.

For the critical exponent $\delta$, the analytical analysis are almost the same. One can assume
\begin{eqnarray}
\chi=\chi_{T_c,1}\delta m^{\kappa_1}+\chi_{T_c,2} \delta m^{\kappa_2}+\chi_{T_c,3}\delta m^{\kappa_3}+....\label{expan-two-m}
\end{eqnarray}
Then the leading expansion are almost the same as Eq.(\ref{eq-chi0}), replacing $\chi_{m_c,1}$ with $\chi_{T_c,1}$:
\begin{eqnarray}\label{eq-chim0}
\ddot{\chi}_{T_c,1}-(\frac{3}{s}+\frac{4s^3}{1-s^4}+\dot{\phi}_{T_c})\dot{\chi}_{T_c,1}+ \frac{3}{s^2(1-s^4)}\chi_{T_c,1}=0.
\end{eqnarray}
Thus, the solutions are exactly the same as Eq.(\ref{eq-chi0}). The non-singular solution $\chi_{1}$ satisfies $\dot{\chi}_{1}(s=0)=0$ and $\chi_1(1)=1$. For the next to leading expansion, there are slightly differences. Since we fix the temperature to $T_c$, there is no next to leading expansion of dilaton field $\phi_T(s)$. Assuming $\kappa_2=\kappa_1 (n-1)$, one gets
\begin{eqnarray}\label{eq-chi1}
\ddot{\chi}_{T_c,2}-(\frac{3}{s}+\frac{4s^3}{1-s^4}+\dot{\phi}_{T_c}))\dot{\chi}_{T_c,2}+ \frac{3}{s^2(1-s^4)}\chi_{T_c,2}= \frac{n\lambda\chi_{m,1}^{n-1}}{s^2(1-s^4)}.
\end{eqnarray}
Since there is no cancellation from dilaton expansion term $g_T$, only the regularity condition $0<|\chi_{m,2}|<\infty$ could be satisfied. $\dot{\chi}_{m,2}(s=0)$ could not be zero. But because we are tuning quark mass away from the critical value $m=0$, the boundary condition of $\chi_{m,2}$ would be different. In the $z$ coordinate, we have $\chi(z)=m_q \zeta z+o(z^2)$, which gives the $s$ coordinate expansion of $\delta\chi=\delta m \zeta z_h s$ when tuning $m_q$ from $0$ to $\delta m$. This would require $\kappa_2=1$. Considering all the above, one reaches $\kappa_1=\frac{\kappa_2}{n-1}=\frac{1}{n-1}$ and $\delta\chi=\chi_{m,1}\delta m^{\frac{1}{n-1}}+...$. Considering the relationship between $\chi_1$ and $\sigma$, one gets $\delta=n-1$.

From the above analysis, one gets $\beta=\frac{\gamma}{n-1}$ and $\delta=n-1$ analytically. From this result, we could see that both the profile of $\Phi$ and $V(\chi)$ are important for the critical behavior of chiral condensate. If dilaton field stands for the affect from gluon sector and the scalar potential stands for the effective interaction in flavor sector, it is reasonable to see that both the two interactions are important for the critical behavior of chiral condensate. Now we consider the more realistic model, the dilaton field in Eq.(\ref{int-dilaton}) and $V(\chi)=-\frac{3}{2}\chi^2+v_4 \chi^4$. Then we have $\gamma=1, n=4$, which gives $\beta=\frac{1}{2}, \delta=3$, consistent with our numerical calculation. In Sec.\ref{sec-twof}, we also take several different $\gamma,n$ groups to test this analytical formula. There, we could see that the numerical results agree with the analytical analysis.

\subsection{Expansion around $\chi\neq0$}
\label{ana-three}

To reduce numerical errors, we also try to extract the critical exponents for $N_f=3$ analytically. The procedure is almost the same as in two-flavor case, except that the critical solution at $T_c$ is not zero. For $\beta$, we begin with the expansion of $\chi$ as
\begin{eqnarray}
\chi=\chi_{m_c,0}+\chi_{m_c,1}\delta T^{\epsilon_1}+\chi_{m_c,2}\delta T^{\epsilon_2}+\chi_{m_c,3} \delta T^{\epsilon_3}+...,
\end{eqnarray}
with $0<\epsilon_1<\epsilon_2<\epsilon_3$. Inserting this expansion into Eq.(\ref{eom-chi3-s}) and keeping the leading order, we get
\begin{eqnarray}
\ddot{\chi}_{m_c,0}(s)-(\frac{3}{s}+\frac{4s^3}{1-s^4}+\dot{\phi}_T)\dot{\chi}_{m_c,0}- \frac{1}{s^2(1-s^4)}(3-3v_3\chi_{m_c,0}^2-4v_4\chi_{m_c,0}^3)=0,\label{eom-chi30-s}
\end{eqnarray}
which is just the rewriting of Eq.(\ref{eom-chi3-s}) and still a non-linear equation. In principle, we should expand the solution at $T_c$. However, at this stage, we can consider the expansion around a general temperature $T$. Actually, we have already solved $\chi_{m_c,0}$ when getting Fig.\ref{sigma-T-degenerate}(b). Since in $z$ coordinate $\chi(z)\approx m_q z\sim \frac{m_q s}{\pi T}$, it is easy to have $\delta\dot{ \chi}(0)\propto \delta T$. Thus, for any power $\epsilon_i$ smaller than $1$, the corresponding $\chi_{m_c,i}$ should satisfy $\dot{\chi}_{m_c,i}(0)=0$. Having this condition, we can continue to analyze the next to leading order solution. There are several different situations. Firstly, if $\epsilon_1=1$, one gets the next to leading order expansion as
\begin{eqnarray}\label{eq-chi3-1}
\ddot{\chi}_{m_c,1}-(\frac{3}{s}+\frac{4s^3}{1-s^4}+\dot{\phi}_{T_c})\dot{\chi}_{m_c,1}+ \frac{3-6v_3\chi_{m_c,0}-12v_4\chi_{m_c,0}^2}{s^2(1-s^4)}\chi_{m_c,1}=-g_T \dot{\chi}_{m_c,0}.
\end{eqnarray}
Generally, the solution of $\chi_{m_c,1}$ in the above equation satisfies $\dot{\chi}_{m_c,1}(0)\neq0$, so the boundary condition $\delta\dot{\chi}(0)\propto \delta T$ could be satisfied. In this case, the expansion power of $\sigma$ is just $\epsilon_1=1$, which is just the normal case as the blue rectangle dot  in Fig.(\ref{sigma-T-degenerate})(b). However, at certain point, this condition could not be satisfied together with $0<|\chi_{m_c,1}(1)|<\infty$. In this case, in order to make sure $\delta\dot{\chi}(0)\propto \delta T$, it requires $2\epsilon_1=\epsilon_2=1$, and the next to leading order and the next next to leading order expansion become
\begin{eqnarray}\label{eq-chi3-2}
\ddot{\chi}_{m_c,1}-(\frac{3}{s}+\frac{4s^3}{1-s^4}+\dot{\phi}_{T_c})\dot{\chi}_{m_c,1}+ \frac{3-6v_3\chi_{m_c,0}-12v_4\chi_{m_c,0}^2}{s^2(1-s^4)}\chi_{m_c,1}=0&&,\\
\ddot{\chi}_{m_c,2}-(\frac{3}{s}+\frac{4s^3}{1-s^4}+\dot{\phi}_{T_c})\dot{\chi_2}_{m_c,2}+ \frac{3-6v_3\chi_{m_c,0}-12v_4\chi_{m_c,0}^2}{s^2(1-s^4)}\chi_{m_c,1}&&\nonumber\\
=-g_T \dot{\chi}_{m_c,0}+\frac{3v_3+12v_4\chi_{m_c,0}}{s^2(1-s^4)}\chi_{m_c,1}^2&&.
\end{eqnarray}
Actually, this is the situation of the blue triangle dot in Fig.\ref{sigma-T-degenerate}(b). However, when quark mass changes, this condition can not be satisfied again. Then, it requires $3\epsilon_1=\epsilon_1+\epsilon_2=\epsilon_3=1$, and the corresponding expansion become
\begin{eqnarray}\label{eq-chi3-3}
\ddot{\chi}_{m_c,1}-(\frac{3}{s}+\frac{4s^3}{1-s^4}+\dot{\phi}_{T_c})\dot{\chi}_{m_c,1}+ \frac{3-6v_3\chi_{m_c,0}-12v_4\chi_{m_c,0}^2}{s^2(1-s^4)}\chi_{m_c,1}=0&&,\\
\ddot{\chi}_{m_c,2}-(\frac{3}{s}+\frac{4s^3}{1-s^4}+\dot{\phi}_{T_c})\dot{\chi}_{m_c,2}+ \frac{3-6v_3\chi_{m_c,0}-12v_4\chi_{m_c,0}^2}{s^2(1-s^4)}\chi_{m_c,2}&&\nonumber\\
=\frac{3v_3+12v_4\chi_{m_c,0}}{s^2(1-s^4)}\chi_{m_c,1}^2&&,\\
\ddot{\chi}_{m_c,3}-(\frac{3}{s}+\frac{4s^3}{1-s^4}+\dot{\phi}_{T_c}(s))\dot{\chi}_{m_c,3}+ \frac{3-6v_3\chi_{m_c,0}-12v_4\chi_{m_c,0}^2}{s^2(1-s^4)}\chi_{m_c,3}(s)&&\nonumber\\
=-g_T \dot{\chi}_{m_c,0}+\frac{6v_3+24v_4\chi_{m_c,0}}{s^2(1-s^4)}\chi_{m_c,1}\chi_{m_c,2}+\frac{4v_4}{s^2(1-s^4)}\chi_{m_c,1}^3&&.
\end{eqnarray}
This is just the situation of the critical point in Fig.\ref{sigma-T-degenerate}(b). Thus, we get the leading expansion of $\chi$ around the critical point as $\chi\simeq \chi_{m_c,0}+\chi_{m_c,1}\delta T^{\tau_1} $, with $\epsilon_1=\frac{1}{3}$. According to the expansion of $\chi_{m_c,1}$, we have $\sigma-\sigma_c\simeq \delta T^{\frac{1}{3}}$, which gives $\beta=\frac{1}{3}$. It is quite easy to understand that the analysis on critical exponent $\delta$ are almost the same, except that we do not have to expand $\phi_T(s)$, which does not depend on quark mass in current settings. Though, in a full back-reaction model, it should depend on quark mass. Therefore, it is easy to get $\delta=3$. The expansion for $N_f=2+1$ in the blue segment is similar. The only difference is that we should expand the coupled equations, which are more complicated. Since the method is the same, we would not repeat it here. For the tri-critical point, since it is quite complicated to expand to very high order, we only do the numerical analysis.

\end{document}